\documentclass[onecolumn,a4paper,amsmath,amssymb,floatfix,showpacs]{revtex4-2}
\usepackage{braket}
\usepackage{bbm}     
\usepackage{graphicx}
\usepackage{dcolumn}
\usepackage{epstopdf}
\usepackage[colorlinks=false,hidelinks]{hyperref}
\usepackage[caption=false]{subfig}

\renewcommand{\Re}{\mathfrak{Re}}
\renewcommand{\Im}{\mathfrak{Im}}
\newcommand{\Op}[1]{\boldsymbol{\mathsf{\hat{#1}}}}

\newcommand{\Fkt}[1]{\,\mathsf {#1}}

\newcommand{\dd}{\ensuremath{\, \textnormal{d}}} 
\newcommand{\old}{\operatorname{old}}
\newcommand{\new}{\operatorname{new}}
\newcommand{\avg}{\operatorname{avg}}
\usepackage{textcomp} 
\newcommand{\micro}{\textmu}
\newcommand{\sqrtISWAP}{\ensuremath{\sqrt{i\text{SWAP}}}}
\newcommand{\identity}{\mathbbm{1}}
\newcommand{\vectorize}{\operatorname{vec}}
\def\mat#1{\hat{#1}}
\def\half{ \frac{1}{2}}
\def\openone{\leavevmode\hbox{\small1\kern-3.3pt\normalsize1}}
\ifx\Tr\renewcommand{\Tr}{\Fkt{Tr}}
\else\newcommand{\Tr}{\Fkt{Tr}}
\fi

\begin{document}

\title{Optimal control theory for a unitary operation under
  dissipative  evolution}

\author{Michael H. Goerz}
\affiliation{U.S. Army Research Lab, Computational and Information Science Directorate, 2800 Powder Mill Rd, Adelphi, MD 20783, USA}

\author{Daniel M. Reich}
\affiliation{Theoretische Physik, Universität Kassel, Heinrich-Plett-Str. 40, D-34132 Kassel, Germany}
\affiliation{Dahlem Center for Complex Quantum Systems and Fachbereich Physik, Freie Universität Berlin, Arnimallee 14, D-14195 Berlin, Germany}

\author{Christiane P. Koch}
\affiliation{Theoretische Physik, Universität Kassel, Heinrich-Plett-Str. 40, D-34132 Kassel, Germany}
\affiliation{Dahlem Center for Complex Quantum Systems and Fachbereich Physik, Freie Universität Berlin, Arnimallee 14, D-14195 Berlin, Germany}

\date{\today}

\begin{abstract}
  We show that optimizing a quantum gate for
  an open quantum system requires the time evolution
  of only three states irrespective of the dimension of Hilbert space.
  This represents a significant reduction in computational resources
  compared to the complete basis of Liouville space that is commonly
  believed necessary for this task. The reduction is based on two
  observations: The target is not a general dynamical map but a unitary
  operation;
  and the time evolution of two properly chosen states is sufficient
  to distinguish any two unitaries.
  We illustrate gate optimization employing a reduced set of states  for a
  controlled phasegate with trapped atoms
  as qubit carriers and a \sqrtISWAP{} gate with superconducting
  qubits.
\end{abstract}

\maketitle

\section{Introduction}
\label{sec:intro}

Quantum effects such as entanglement and matter interference are
predicted cornerstones of future technologies. Their exploitation
requires the ability to reliably and accurately control complex
quantum systems. A major obstacle is that a quantum system
can never completely be isolated from its environment and the
interaction with the environment causes
decoherence~\cite{BreuerBook}. This is particularly true for condensed
phase settings as encountered in, e.g., solid state quantum devices.
A number of concepts, such as decoherence-free subspaces~\cite{LidarPRL98}
and noiseless subsystems~\cite{LorenzaSci01}, dynamical
decoupling~\cite{LorenzaPRL99} and spectral
engineering~\cite{ClausenPRL10}, have been developed to cope with
decoherence. The applicability of these strategies is tied to specific
conditions on the interaction between system and environment and,
in practice, is often  limited to systems that can be described by simple
models. For complex quantum systems, numerical optimal control offers an
alternative approach. It calculates the external controls that
implement a desired target operation by performing an iterative search
in the parameter space of the controls~\cite{RiceBook}.

For quantum systems that are subject to decoherence, numerical optimal
control was first employed
to realize laser cooling of internal degrees of freedom in
molecules~\cite{BartanaJCP97}. Further applications, also utilizing a
Markovian master equation to describe the open system dynamics,
include controlling coherences~\cite{OhtsukiJCP99}, automatic
protection against noise~\cite{KallushPRA06}, selective
photoexcitation of charge transfer~\cite{TremblayPRA08},
electric current in a molecular junction~\cite{KleinekathoeferEPJB10},
quantum gates~\cite{ToSHJPB11}, and
quantum memories~\cite{GormanPRA12}. Due to the formal equivalence
between Markovian
dissipation and quantum measurements, optimized observations can be
determined using the same set of tools~\cite{ShuangJCP07}.
Numerical optimal control can also be applied to non-Markovian
quantum
systems~\cite{RebentrostPRL09,AsplundPRL11,SchmidtPRL11,FloetherNJP12}
provided the dynamics can be calculated with sufficient efficiency.

The question of numerical effort becomes particularly important in the
optimization of high-fidelity quantum gates. High
fidelities, or small errors,  are best achieved with monotonically convergent
optimization algorithms that utilize
gradient information and thus require repeated forward and backward
propagation~\cite{SomloiCP93,ZhuJCP98}. Gate optimization under
coherent dynamics implies propagation of a set of states that span the
Hilbert (sub)space on which the target is
defined~\cite{JosePRL02,JosePRA03}. For open system dynamics, this was
generalized to a set of states that span the corresponding Liouville
(sub)space~\cite{KallushPRA06,OhtsukiNJP10,ToSHJPB11,FloetherNJP12}. It
requires not only propagation of density matrices instead of
wavefunctions but also a significantly larger number of states
since Liouville space dimension is the square of Hilbert space
dimension. Realistically, this limits quantum gate optimization to but
the simplest examples, i.e., one-qubit and two-qubit operations.

The direct extension from Hilbert to
Liouville space~\cite{KallushPRA06,OhtsukiNJP10,ToSHJPB11}
overlooks the fact that in quantum gate optimization,
the target is a unitary operation and not a general dynamical map. The
latter would indeed require a basis that spans the full Liouville
space. However, much less information is required to assess how
well a desired unitary is implemented. This observation is not only
relevant for optimal control but also provides the
basis for all current attempts at reducing the resources for
estimating the average gate
error~\cite{BenderskyPRL08,FlammiaPRL11,MagesanPRL11,daSilvaPRL11,ReichKochPRL13}.
In fact, only two states are necessary to distinguish any two
unitaries, irrespective of Hilbert space
dimension~\cite{ReichKochPRA13}.
We show here that these two states,
together with a third state enforcing the dynamical map on the
optimization subspace to be
contracting and population conserving,
can be utilized to construct an optimization functional
which attains its optimal value only if the desired gate is
implemented with unit fidelity.

The two states that are required for unitary identification are
constructed such that the first one consists of non-degenerate
contributions from each Hilbert (sub)space direction. This corresponds
to choosing a basis, and probing the gate error within this basis. In
order to determine the error of gates that are diagonal in the chosen
basis, i.e., phase errors, the second state is needed. For Hamiltonians
which due to their inherent structure allow for nothing but diagonal gates,
only the second state together with the third one is required, enforcing the
dynamical map on the optimization subspace to be contracting and
norm conserving. In our application, we
thus distinguish between gates which are diagonal and those that are
non-diagonal in the logical basis.

Our optimization functional is closely related to the gate error.
While two states represent the minimal set of states required to
distinguish any two unitaries, it is impossible to deduce bounds on
the gate error from the two
states~\cite{ReichKochPRA13}. This is due to the state corresponding to
the choice of basis being a totally mixed thermal state. Meaningful
bounds on the gate error can be derived numerically when replacing the totally
mixed state by a set of $d$ pure states where $d$ is the dimension of
Hilbert space, i.e., by choosing a separate basis state for each
Hilbert space direction~\cite{ReichKochPRA13,FiurasekPRA14}. The resulting set
consists of $d+1$ states. Analytical bounds are obtained when also the
second state of the minimal set is expanded~\cite{HofmannPRL05}. The
corresponding set is built out of the $2d$ states of two mutually
unbiased bases~\cite{ReichKochPRA13}. This observation from process
verification motivates the choice of optimization functionals which
utilize these extended sets of states. Although the
number of states then depends on Hilbert space dimension, this choice still
comes with very significant savings in the computational resources.
For example, already for a two-qubit gate,
both $2d$ and $d+1$ represent a significant reduction in the number of
states that need to be propagated, namely a reduction from 16 for the
full Liouville space basis to 8 and 5, respectively.

We demonstrate below that two states are sufficient to
optimize diagonal gates and three states to optimize non-diagonal
two-qubit gates. We also show that, depending on the desired gate
error, $d+1$, respectively $2d$ states in the optimization functional
correspond to the numerically most efficient choice. We consider a
controlled phasegate  with neutral trapped atoms that are excited into
a Rydberg state and a \sqrtISWAP gate with
superconducting qubits. In both examples, our optimization identifies
gate implementations for which the error is limited by
decoherence. This proves that all reduced sets of states are
sufficient for determining the fundamental limit to the gate error and
thus for quantum gate optimization.

The paper is organized as follows. Section~\ref{sec:oct} defines
the optimization functional and presents the optimization
algorithm. Optimization of a controlled phasegate for neutral atoms is
discussed in Sec.~\ref{sec:phasegate}, whereas optimization of a
non-diagonal gate for superconducting qubits is studied in
Sec.~\ref{sec:allgates}. Section~\ref{sec:concl}
concludes.
The algebraic framework and the proofs required for the construction
of the three states employed in the optimization functional are
presented in Appendix~\ref{sec:proof}.

\section{Optimal control theory for a unitary operation under
  dissipative  evolution}
\label{sec:oct}

\subsection{Optimization functional}
\label{subsec:func}
In order to employ optimal control theory to determine a
high-fidelity implementation of quantum gates, one needs to
define a distance measure $J_T $
between the desired unitary $\Op O$ and the actual evolution. We show
here that
\begin{equation}
  J_T = 1 - \sum_{i=1}^{n}
    \frac{w_i}{\Tr[\Op\rho_i^2(0)]} \, \mathfrak{Re}\left\{\Tr\left[
    \Op O\Op\rho_{i}(0)\Op O^{\dagger}\Op\rho_{i}\left(T\right)\right]\right\}
  \label{eq:functional}
\end{equation}
with $n=3$ and specific initial states $\Op\rho_i(0)$ represents
a suitable choice for $J_T$. This is in contrast to
Refs.~\cite{KallushPRA06,OhtsukiNJP10,ToSHJPB11}, where $n$ was taken
to be the Liouville space dimension corresponding to $\Op O$,
i.e., $n=2^{2N}$ for $N$ qubits,
and $\Op\rho_i$ an orthonormal basis (under the Hilbert-Schmidt
product) of Liouville space.
In Eq.~\eqref{eq:functional}, $w_i$ are  weights,
normalized as $\sum_{i=1}^n w_i = 1$. In order to evaluate $J_T$,
the time evolved states $\Op\rho_i(T)$
need to be obtained by solving the equation of motion describing the open
system's evolution for $\Op\rho_i$. While in general the dynamics can
be non-Markovian, we will restrict ourselves to a Markovian master
equation in the examples below. We assume the coherent part to include
coupling to
an external control, i.e., the Hamiltonian is of the form $\Op
H(t)=\Op H_0 + \epsilon(t)\Op H_1$, and generalization to several
controls $\epsilon_i(t)$ is straightforward.

The functional $J_T$ needs to be minimized
with respect to $\epsilon(t)$.
Further constraints can be added, for example,
\begin{equation}
  \label{eq:J}
  J = J_T -
  \lambda_a\int_0^T\left[\epsilon(t)-\epsilon_{\text{ref}}(t)\right]^2/S(t) \dd t\,,
\end{equation}
where $\epsilon_{\text{ref}}(t)$ denotes a reference field, $S(t)$ enforces
the field to be smoothly switched on and off,
and the second term in Eq.~\eqref{eq:J} ensures a finite pulse
fluence~\cite{JosePRA03}. More complex additional constraints, for
example restricting the spectral width of the pulse or confining the
accessible state space~\cite{ReichKochJMO13,JosePRA13}, are also
conceivable.

Mathematically, our claim  that only three states are sufficient to
determine proper implementation of the desired unitary $\Op O$
is equivalent to the conjecture that the optimization functional
attains its global minimum if and only if
\begin{equation}
  \label{eq:cond}
  \Op\rho_{i}\left(T\right) =
  \Op O\Op\rho_{i}(0)\Op O^{\dagger}
\end{equation}
for the three states $\Op\rho_i$. The three states are constructed
such that the first one fixes a basis, and the corresponding
Hilbert-Schmidt product in Eq.~\eqref{eq:functional} checks whether
the gate is correctly implemented in this basis. It misses errors
for gates that are diagonal in the basis, i.e., phase
errors~\cite{ReichKochPRA13}. The second state is therefore chosen to
detect phase errors with its Hilbert-Schmidt product in
Eq.~\eqref{eq:functional}~\cite{ReichKochPRA13}. The Hilbert-Schmidt
product of the third state determines
whether the dynamical map attained at time $T$ conserves the
population within the optimization subspace. This is necessary since
the time evolution can  be non-unitary due to decoherence or due to
leakage into states other than the logical basis~\footnote{%
  Strictly speaking, one should enforce the dynamical map on the
  optimization subspace to be unital, i.e., both norm conserving
  and contracting. This could only be achieved by employing the trace
  distance  of the ideal and actual time evolved third state,
  not their Hilbert Schmidt product, in the optimization functional,
  cf. Appendix~\ref{sec:proof}. However, for all practical purposes,
  the Hilbert Schmidt product turns out to be sufficient.
}.

In more technical terms, $\Op\rho_{1}(0)$ is a density matrix with $N$
non-degenerate,  non-zero eigenvalues.
Spanning the $d$-dimensional Hilbert space ($d=2^N$ for $N$ qubits)
by an arbitrary complete
orthonormal basis, $\{|\varphi_i\rangle\}$,
$\Op\rho_{1}(0)$ is expressed in terms of a complete
set of $d$ one-dimensional orthogonal projectors
$\Op P_i=|\varphi_i\rangle\langle\varphi_i|$, i.e.,
$\Op\rho_{1}(0) = \sum_{i=1}^d\lambda_i \Op P_i$ with
$\lambda_i\neq\lambda_j\forall i\neq j$ and $\lambda_i\ge
0$~\cite{ReichKochPRA13}. The
second state, $\Op\rho_{2}(0)$, is constructed to be
totally rotated with respect to $\Op\rho_{1}(0)$, i.e.,
$\Op\rho_{2}(0)=\Op P_{TR}$ where $\Op P_{TR}$ is a
one-dimensional projector obeying $\Op P_{TR}\Op P_i\neq 0$ for
$i=1,\ldots,d$~\cite{ReichKochPRA13}.
$\Op\rho_{3}(0)$ is the identity in the optimization subspace. A
possible choice for the initial states reads
\begin{subequations}\label{eq:rhos}
  \begin{eqnarray}
    \left(\Op\rho_{1}(0)\right)_{ij} & = &
    \frac{2\left(d-i+1\right)}{d\left(d+1\right)}\delta_{ij}\,,\label{eq:rho1}\\
    \left(\Op\rho_{2}(0)\right)_{ij} &=& \frac{1}{d}\,,\label{eq:rho2}\\
    \left(\Op\rho_{3}(0)\right)_{ij} &=& \frac{1}{d}\delta_{ij}\,,\label{eq:rho3}
  \end{eqnarray}
\end{subequations}
where the matrix elements are given in the optimization subspace,
all other elements are zero.
We show in Appendix~\ref{sec:proof} that the optimization reaches its target
if and only if condition~\eqref{eq:cond} is fulfilled. Specifically, we
prove that propagation of three states is sufficient, irrespective of
the dimension of the optimization subspace. Already for a small number
of qubits, this represents a
significant computational saving compared to the propagation of $2^{2N}$
initial states deemed necessary in the
literature~\cite{KallushPRA06,OhtsukiNJP10,ToSHJPB11}.

The states $\Op\rho_1$ and $\Op\rho_2$ of Eq.~\eqref{eq:rhos}, while
sufficient in principle to distinguish any two unitaries, do not allow
for stating bounds on the gate
error~\cite{ReichKochPRA13}. Meangingful bounds on the gate error can
be obtained numerically by replacing $\Op\rho_1$,
$\Op\rho_2$ by a set of $d+1$ states, whereas analytical
bounds can be deduced from  $2d$
states~\cite{ReichKochPRA13,HofmannPRL05,FiurasekPRA14}. Motivated by this fact,
we define two additional sets of states that can be employed in
Eq.~\eqref{eq:functional}. When $n$ in Eq.~\eqref{eq:functional} is
taken to be  equal to $d+1$, the totally mixed state of
Eq.~\eqref{eq:rho1} is replaced by $d$ pure states,
\begin{equation}
  \label{eq:rho1_dp1}
  \Op\rho_j(0) = |\varphi_j\rangle\langle\varphi_j|\,,
\end{equation}
with $j=1,\ldots,d$ and $\{|\varphi_j\rangle\}$ the logical basis.
$\Op\rho_{d+1}(0)$ is simply equal to $\Op\rho_2(0)$ of
Eq.~\eqref{eq:rho2}. In this case, Eq.~\eqref{eq:rho3} is not required since
the $d+1$ pure states are sufficient to enforce the dynamical map on
the optimization subspace to be contracting and norm conserving.
Similarly, the functional~\eqref{eq:functional} employing
$n=2d$ states is constructed by replacing $\Op\rho_1(0)$ of
Eq.~\eqref{eq:rho1} by $\Op\rho_j$, $j=1,\ldots,d$ of
Eq.~\eqref{eq:rho1_dp1} and $\Op\rho_2(0)$ of
Eq.~\eqref{eq:rho2} by
\begin{equation}
  \label{eq:rho2_2d}
  \Op\rho_{d+j}(0) = |\tilde\varphi_{j}\rangle\langle\tilde\varphi_{j}|\,,
\end{equation}
with $j = 1,\dots,d$,
where the states $\Ket{\tilde\varphi_j}$ form a mutually unbiased basis with
respect to the canonical basis $\{\Ket{\varphi_j}\}$. For two
qubits  ($d=4$), an example for such a basis is given by
\begin{subequations}\label{eq:mub}
  \begin{eqnarray}
    \Ket{\tilde\varphi_{1}}
    &=& \frac{1}{2} \left( \Ket{00} + \Ket{01} + \Ket{10} + \Ket{11} \right) \,,\\
    \Ket{\tilde\varphi_{2}}
    &=& \frac{1}{2} \left( \Ket{00} - \Ket{01} + \Ket{10} - \Ket{11} \right) \,,\\
    \Ket{\tilde\varphi_{3}}
    &=& \frac{1}{2} \left( \Ket{00} + \Ket{01} - \Ket{10} - \Ket{11} \right) \,,\\
    \Ket{\tilde\varphi_{4}}
    &=& \frac{1}{2} \left( \Ket{00} - \Ket{01} - \Ket{10} + \Ket{11} \right) \,.
  \end{eqnarray}
\end{subequations}

\subsection{Optimization algorithm}
\label{subsec:krotov}

We assume in the following a coupling to the external field that is linear
in the field and equations of motion that are linear in the
states~\footnote{%
  A generalization to non-linear couplings and equations of motion is
  straightforward following Ref.~\cite{ReichKochJCP12}.}.
Moreover, the full optimization functional,
Eq.~\eqref{eq:J}, is linear in the states $\Op\rho_i(T)$ and does
not depend on the states at intermediate times $t$. In this case, the
linear version of Krotov's method is sufficient to yield a
monotonically convergent optimization
algorithm~\cite{ReichKochJCP12}. It is given in terms of coupled
control equations that need to be solved simultaneously.
Here, we model the dissipative time evolution by a
Markovian master equation,
\begin{equation}
  \label{eq:LvN}
  \frac{d\Op\rho}{dt} = \mathcal{L}(\Op\rho) = -i[\Op H(t),\Op\rho]
  +\mathcal{L}_D(\Op\rho)\,.
\end{equation}
The control equations then read
\begin{subequations}\label{eq:control}
  \begin{eqnarray}
    \label{eq:forward}
    \frac{d\Op\rho_i}{dt} &=& -i[\Op H,\Op\rho_i] +\mathcal{L}_D(\Op\rho_i)\,,\\
    \label{eq:backward}
    \frac{d\Op\sigma_i}{dt}
    &=& -i[\Op H,\Op\sigma_i] -\mathcal{L}^{\dagger}_D(\Op\sigma_i)
    \quad\mathrm{and}\quad
    \Op\sigma_i(t=T) =
    \frac{w_i}{\Tr[\Op\rho_i^2(0)]}
     \Op O \Op \rho_i(0) \Op O^{\dagger}\,,\\
     \label{eq:update}
     \Delta\epsilon(t) &=&
     \frac{S(t)}{\lambda_a} \sum_{i=1}^n \Im\left\{
     \Tr\left(
       \Op\sigma_i^{\old}(t)
       \frac{\partial \mathcal{L}\left(\Op\rho_i\right)}{\partial \epsilon}
       \Big|_{\rho_i^\mathrm{new},\epsilon^\mathrm{new}}
     \right)\right\}
   \end{eqnarray}
\end{subequations}
with $i=1,2,3$ when the initial conditions $\Op\rho_i(0)$
of Eq.~\eqref{eq:rhos} are
employed or $i=1,\ldots d^2$ with $d$ the dimension of Hilbert space
when a full basis of Liouville space is propagated.
Eq.~\eqref{eq:backward} is derived in full detail in Appendix~\ref{sec:erratum}.
In Eq.~\eqref{eq:update},
the states $\Op\sigma^{\old}_i$ are backward-propagated with the
pulse of the previous iteration ('old'), whereas the states
$\Op\rho^{\new}_i$ are forward-propagated with the updated
pulse ('new'). The derivative with respect to the field is given by
the commutator
\begin{equation}
\frac{\partial \mathcal{L}\left(\Op\rho\right)}{\partial \epsilon}
= -i \left[\frac{\partial \Op H}{\partial \epsilon}, \Op{\rho} \right]
\end{equation}
and has to be evaluated for the `new' field and the states $\Op\rho$ propagated
under the `new' field. For a complex control, which
occurs for example when using the rotating wave approximation (RWA),
Eq.~\eqref{eq:update} holds for both the real and the imaginary part
of   $\epsilon(t)$.

The value of the optimization functional in Eq.~(\ref{eq:functional}) depends on
the number and the specific choice of initial states as well as the choice of
weights. It is therefore not suitable to compare the convergence behavior
between different sets of states. Instead, we employ the average gate
fidelity,
\begin{equation}
  F_{\avg} = \int \langle \Psi | \Op{O}^{\dagger}
              \mathcal{D}(\Ket{\Psi}\!\Bra{\Psi})
             \Op{O} | \Psi \rangle \dd \Psi\,,
  \label{eq:Favg}
\end{equation}
for the comparison.
In Eq.~\eqref{eq:Favg}, $\mathcal D$ denotes the dynamical map
describing the time evolution of the open quantum system, i.e.,
$\Op\rho(T)=\mathcal D\left(\Op\rho(0)\right)$.
The gate fidelity, respectively the gate error, $1 - F_{\avg}$, is easily
evaluated as~\cite{PedersenPLA07}
\begin{equation}
  F_{\avg} = \frac{1}{d (d+1)} \sum_{i,j=1}^d \left(
              \langle \varphi_i |
                \Op{O}^\dagger
                \mathcal{D}(\Ket{\varphi_i}\!\Bra{\varphi_j})
                \Op{O} |
              \varphi_j \rangle
              + \Tr\left[
                \Op{O}\Ket{\varphi_i}\!\Bra{\varphi_i}\Op{O}^{\dagger}
                \mathcal{D}(\Ket{\varphi_j}\!\Bra{\varphi_j})
              \right]
           \right)\,.
\end{equation}

\section{Example I: Diagonal gates}
\label{sec:phasegate}

\begin{figure}[tb] 
  \centering
  \includegraphics{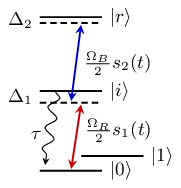}
  \caption{Atomic levels for two-photon near-resonant excitation to a
    Rydberg state.}
  \label{fig:levels}
\end{figure}
It is quite common that a two-qubit
Hamiltonian allows only for diagonal gates, such as a controlled
phasegate. A prominent example are non-interacting
qubit carriers that interact only when excited into an auxiliary state
where they accumulate a non-local phase~\cite{JakschPRL00}.
Neutral trapped atoms with long-range interaction in a Rydberg
state present a physical implementation of this
setting~\cite{JakschPRL00,SaffmanRMP10}.
Optimal control theory has been employed before to determine the
minimum time in which a controlled phasegate can be
implemented~\cite{GoerzKochJPB11} and the optimum distribution of the
single-qubit phases~\cite{MuellerKochPRA11}. These optimizations were
carried out, however, without explicitly accounting for decoherence.
It is thus not clear whether the best solutions to avoid decoherence
have indeed been identified.
While the logical basis states and the Rydberg state are typically
very long-lived, the main source of decoherence is spontaneous decay
from an intermediate state which is necessary to access the Rydberg
state.
Due to experimental feasibility, the excitation to the Rydberg state
proceeds by a near-resonant two-photon process.
The corresponding single atom Hamiltonian in the basis
$\{\Ket{0},\Ket{1},\Ket{i},\Ket{r}\}$, cf. Fig.~\ref{fig:levels}, and
employing a two-color rotating wave approximation is given by
\begin{subequations}
  \label{eq:H_Ryd}
\begin{equation}
  \label{eq:H_1atom}
  \Op{H}_{1q} =
  \begin{pmatrix}
    0  & 0  & \frac{1}{2}\Omega_R(t) & 0                    \\
    0  & E_1 & 0                         & 0                    \\
    \frac{1}{2} \Omega_R(t)  & 0  & \Delta_1 & \frac{1}{2}\Omega_B(t)  \\
       0  & 0  & \frac{1}{2} \Omega_B(t) & \Delta_2
     \end{pmatrix}\,.
\end{equation}
The total Hamiltonian for two atoms includes an
interaction when both atoms are in the Rydberg state,
\begin{equation}
  \label{eq:H_2atom}
  \Op H = \Op H_1\otimes\openone + \openone\otimes\Op H_1
  - U |rr\rangle\langle rr|\,.
\end{equation}
\end{subequations}
Spontaneous emission from the intermediate level is accounted for by
the dissipator
\begin{equation}
  \label{eq:L_atom}
  \mathcal L_D(\Op\rho) = \gamma \left(
    \Op{A} \Op{\rho} \Op{A}^\dagger
    - \frac{1}{2} \left\{\Op{A}^\dagger \Op{A}, \Op{\rho} \right\}
  \right) \quad\mathrm{with}\quad
    \Op{A} = \Ket{0}\Bra{i}\,,
\end{equation}
and $\gamma$ the decay rate, $\gamma=1/\tau$.
The parameters correspond to optically trapped rubidium atoms
and are summarized in Table~\ref{tab:atom}.
Since qubit level $|1\rangle$ remains decoupled throughout the
time evolution, cf. Eq.~\eqref{eq:H_1atom} and Fig.~\ref{fig:levels},
the Hamiltonian~\eqref{eq:H_Ryd} admits only diagonal gates.
\begin{table}[tb]
  \centering
 \begin{tabular}{|l|r|}\hline
  single-photon detuning $\Delta_1$                 & 600~MHz \\ \hline
  two-photon detuning $\Delta_2$                 & 0 \\ \hline
  excitation energy  $E_1$                      & 6.8~GHz \\ \hline
  Rabi frequencies  $\Omega_R$, $\Omega_B$      & 300~MHz \\ \hline
  interaction energy  $U$                        & 50~MHz \\ \hline
  lifetime $\tau = 1/\gamma$ & 25~ns \\\hline
 \end{tabular}
  \caption{Parameters of the Hamiltonian, Eq.~\eqref{eq:H_Ryd},
    for implementing a controlled phasegate with two rubidium
    atoms.}
  \label{tab:atom}
\end{table}
The update equations for real and imaginary part of the red and blue
pulses are obtained by evaluating Eq.~(\ref{eq:update}) for the
Hamiltonian given in Eq.~(\ref{eq:H_Ryd}),
\begin{subequations}\label{eq:control_Ryd}
\begin{eqnarray}
  \label{eq:Re}
  \Re\left\{\Delta \Omega_{R,B}(t)\right\}
  & = &\frac{S(t)}{\lambda_a}
   \sum_{i=1}^{n} \Im\left\{\Tr \left(
      \Op\sigma_{i}^{\old}(t)
      \, \left[\Op{H}_{R,B}(t),
        \Op\rho_{i}^{\new}(t) \right]
    \right)\right\}\\ \label{eq:Im}
  \Im\left\{\Delta \Omega_{R,B}(t)\right\}
  & =& \frac{S(t)}{\lambda_a}
  \sum_{i=1}^{n}   \Im\left\{\Tr \left(
      i \Op\sigma_{i}^{\old}(t)
      \, \left[\Op{H}_{R,B}(t),
        \Op\rho_{i}^{\new}(t)\right]
    \right)\right\}\,,
\end{eqnarray}
\end{subequations}
where $\Op{H}_{R,B}$ represents the
control Hamiltonians coupling to the red and blue laser,
respectively, obtained by rewriting Eq.~(\ref{eq:H_Ryd}) as the
sum of a diagonal drift Hamiltonian and the two control Hamiltonians,
\begin{equation}
\Op{H} = \Op{H}_{\text{drift}}
         + \Omega_R(t) \Op{H}_R  + \Omega_{B}(t) \Op{H}_B\,.
\end{equation}

\begin{figure}[tb] 
  \centering
 \includegraphics{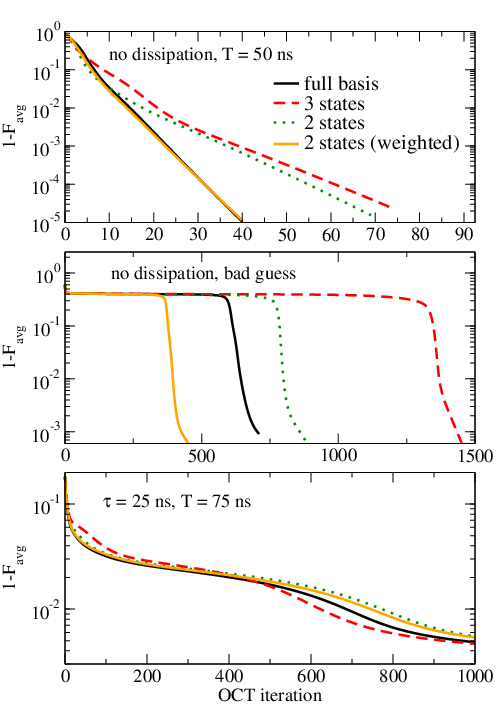}
 \caption{Optimizing a controlled phasegate for two trapped neutral
   atoms that are excited to a Rydberg state. The convergence is shown
   as the gate error, $1-F_{\avg}$, over OCT iterations, using  the full basis of
   16 states (solid black lines), as well as a reduced set of three
   states (red dashed lines) and a reduced set of two states (green
   dotted and orange solid line). The
   calculations employ equal weights of all states, except for those shown in
   orange where $w_2 / w_3 = 10$.  The top and middle panels show
   optimizations without
   any dissipation; the middle panel shows a calculation with the same
   parameters as the top panel except for the guess pulse which is
   badly chosen. The optimization shown in the bottom panel
   takes into account spontaneous emission from the intermediate state, with
   a lifetime of $\tau = 25\,$ns. The gate duration is $T=50\,$ns for the top and
   middle panels, and $T=75\,$ns for the bottom panel. The number of iterations
   and the reached gate error differ significantly in all three
   situations, cf. the different x- and y-axes scales.
 }
  \label{fig:converg:Rydberg}
\end{figure}

Figure~\ref{fig:converg:Rydberg} shows the gate error of the
controlled phasegate versus iteration of the optimization algorithm when
using a full basis, i.e., 16 states, or using three, respectively two,
states in
Eq.~\eqref{eq:control_Ryd}.
The minimum number of states in this
example is two since the Hamiltonian admits only diagonal gates, i.e.,
only phase errors and norm conservation within the logical subspace have to be
checked. Therefore, $\Op\rho_1$ in Eq.~(\ref{eq:rho1}) can be omitted, and the two
remaining states are $\Op\rho_2$ (phase errors) and $\Op\rho_3$ (norm conservation) of
Eqs.~(\ref{eq:rho2},~\ref{eq:rho3}).
The relative weights $w_2$ and $w_3$ in Eq.~\eqref{eq:functional} can
be modified to emphasize one of  the two aspects.
Figure~\ref{fig:converg:Rydberg} therefore also compares two
states with equal and unequal weights in Eq.~\eqref{eq:functional},
cf.\ green dotted and orange solid lines. The fastest convergence was obtained
for $w_2 / w_3 = 10$.
The panels from top to bottom show the optimization without any dissipation,
starting from a well-chosen guess pulse; an optimization starting with a bad
guess pulse of insufficient fluence; and an optimization taking into account
spontaneous decay from the intermediate level.
As the main observation, Fig.~\ref{fig:converg:Rydberg}
clearly demonstrates that only two states are sufficient to optimize a
quantum gate for a Hamiltonian of this kind.
The optimization for coherent time evolution (top panel), shows that
while the use of three states converges to gate errors as small as
those obtained with the full
basis, the convergence rate is only about half that of the full basis.
This is due to two factors: (i) For the optimization with three states,
there is no bound on the distance between the value
$J_T$ and the gate error, such that
the path in the optimization landscape  may be less
direct until an asymptotic value is reached. Since without dissipation, there is
no limit to the gate error, the convergence of $J_T$ and that of
$1-F_{\avg}$ stay on different trajectories.
(ii) The reduced sets of states are constructed specifically
to take into account decoherence. In particular, the third state
contributes significantly less information that is relevant for
reaching the optimization target than the second state.
The convergence can be improved dramatically by weighting the three states
according to the relevance of the information they carry. In this
respect, the use of only two initial
states can be seen as choosing $w_1 = 0$. Taking $w_2 > w_3$
addresses the issue of $\Op\rho_3$ contributing less to the optimization.
Choosing proper weights allows for ensuring the convergence of
optimization with a reduced set of
states to be as fast as the optimization using the full basis.

The importance of choosing weights appropriate to the optimization problem
becomes even more evident when the optimization starts from a bad guess pulse of
insufficient fluence, as shown in the center panel of
Fig.~\ref{fig:converg:Rydberg}. The features observed in
Fig.~\ref{fig:converg:Rydberg} are typical: The
plateau near the beginning corresponds to the optimization increasing
the intensity of the pulse
without any significant improvement in the gate error, before converging
quickly once the pulse is sufficiently intense. The end of the plateau
can be significantly influenced by the
choice of weights, cf.\ solid orange and dotted green curves in the
middle panel of Fig.~\ref{fig:converg:Rydberg}.
Remarkably, the optimal choice of using two properly weighted
initial states outperforms the use of the full basis. This might be explained by the
fact that each of the three states in the reduced set has a specific physical
role to play in the optimization, and this role can be emphasized by choice of
the weight. In contrast, all states in the full basis fulfill the
same role in the optimization, and thus there is no way in which different
weights on individual states would improve the convergence.

One should point out that even in the cases where the use of two or three states
shows a slower convergence than that of the full basis, they still outperform
the full basis in terms of numerical resources. Since both CPU time and the
required memory scale linearly with the number of initial states in the
optimization, using only two states compared to 16 has a 1:8 advantage, which
more than offsets the factor of two in the convergence rate in the
middle panel of Fig.~\ref{fig:converg:Rydberg}.

Naturally, without the presence of decoherence, there is no reason to perform
the optimization in Liouville space. Therefore, the results shown here only
serve to illustrate the general convergence behavior of a reduced set of initial
states. The more relevant case of non-coherent dynamics is shown in the bottom
panel of Fig.~\ref{fig:converg:Rydberg}. The presence of decoherence implies the
existence of an asymptotic bound on the gate error. This
constraint on the optimization landscape (together with the further constraint
that only diagonal gates are reachable) ensures that all sets of reduced states
converge at a similar rate, once the asymptotic region is approached. We expect
that all choices reach the same asymptotic value; which choice yields the best
fidelity after a specific number of iterations cannot be predicted in general.
Factoring in all necessary resources, optimization using two states with unequal
weights dramatically outperforms optimization
using the full basis in this example.

\begin{figure}[tbp] 
  \centering
  \includegraphics{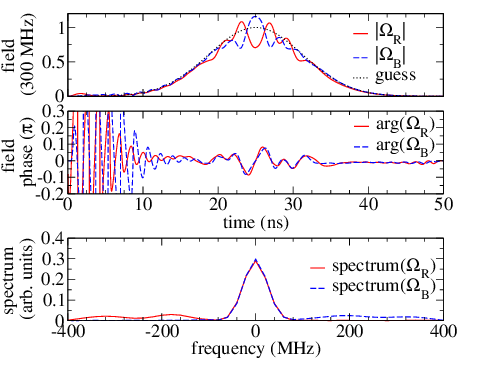}
  \caption{The optimized pulses $\Omega_{B,R}(t)$ for the blue and red laser
    cf.\ Fig~\ref{fig:levels}, resulting from optimization using two
    states with unequal weights without spontaneous decay
    (corresponding to the orange
    solid line in the top panel of Fig.~\ref{fig:converg:Rydberg}). The pulse
    amplitudes are shown in the top panel, the complex phase in the
    center panel, and the pulse spectrum in the bottom panel.  The guess
    pulse, indicated by the black dotted line in the top panel, is
    identical for both the red and the blue laser.  In the spectrum,
    frequency 0 corresponds to the carrier frequencies of the laser pulses.
  }
  \label{fig:pulse:Rydberg:nodiss}
\end{figure}
The optimized pulse and spectrum in the case of coherent dynamics is presented in
Fig.~\ref{fig:pulse:Rydberg:nodiss}. The result shown here is obtained from the
optimization using two initial states with unequal weights. However, the pulse is
indistinguishable from the one obtained using the full basis, consistent with
the identical convergence behavior for the two sets in the upper panel of
Fig.~\ref{fig:converg:Rydberg}. The optimized pulses only show relatively small
amplitude modulations compared to the guess pulse (dotted line). These
modulations appear as small side-peaks in the spectrum. In the time interval in
which there is a significant pulse amplitude, the complex phase only deviates by
about $\frac{\pi}{10}$ from zero. This phase evolution is
reflected in the asymmetry of the spectrum for the red and the blue pulse
(bottom panel). The spectrum nicely illustrates the
mechanism of control: while each spectrum by itself is asymmetric, the
red pulse showing negative frequencies, the blue pulse showing positive
frequencies, the sum of both pulses is again symmetric, i.e., positive and
negative frequencies cancel out. This means that the combination of both pulses
is two-photon resonant with the transition $\Ket{0} \rightarrow \Ket{r}$,
providing multiple pathways for the same transition whose interference might be
exploited by the optimization.

\begin{figure}[tbp] 
  \centering
  \includegraphics{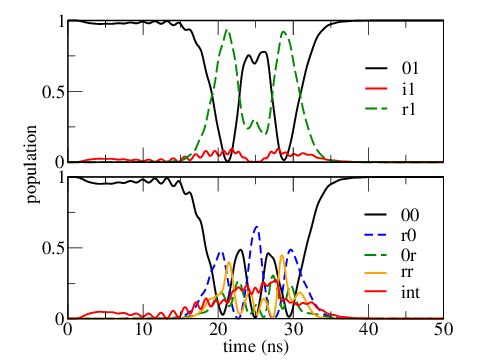}
  \caption{Population dynamics under the pulse shown in
           Fig.~(\ref{fig:pulse:Rydberg:nodiss}), for the logical basis states
           $\Ket{01}$ (top) and $\Ket{00}$ (bottom). The intermediate population
           (''int'') is integrated over all levels with decay, i.e., $\Ket{0i}$,
           $\Ket{i0}$, $\Ket{ii}$, $\Ket{ir}$, and $\Ket{ri}$.
  }
  \label{fig:dyn:Rydberg:nodiss}
\end{figure}
The population dynamics induced by the optimized pulses are shown in
Fig.~\ref{fig:dyn:Rydberg:nodiss}. The two-photon resonance of the pulse
expresses itself in a direct Rabi cycling between $\Ket{0}$ and
$\Ket{r}$ on the left qubit in the propagation of $\Ket{01}$ (top panel). The
population shows roughly a $4\pi$ Rabi flip due to the relatively high pulse
intensity. The nearly 25\% of the population in the intermediate states in the
propagation of $\Ket{00}$ (bottom panel) is due to the fact that the decay from
these levels was not included in the optimization, and thus the optimization
algorithm makes no attempt at suppressing population in these states.

\begin{figure}[tbp] 
  \centering
  \includegraphics{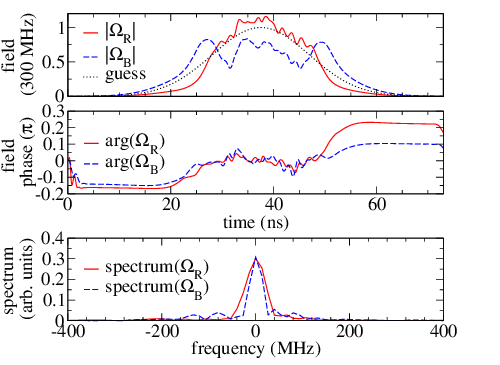}
  \caption{The optimized pulses resulting from optimization using two weighted
           states and including spontaneous decay (orange
           solid line in the bottom panel of Fig.~\ref{fig:converg:Rydberg}), using
           the same conventions as Fig.~\ref{fig:pulse:Rydberg:nodiss}.
  }
  \label{fig:pulse:Rydberg:diss}
\end{figure}
For the optimization with dissipation, the optimized pulse and pulse
spectrum is shown in Fig.~\ref{fig:pulse:Rydberg:diss}. The characteristics of
the pulses are quite different compared to the coherent case. The red pulse
remains close to the single Gaussian peak of the guess pulse, except for
being slightly narrower. The blue pulse has a more complex structure. It is
overall broader than the red pulse and consists of three distinctive features:
an initial peak that overlaps but precedes the red pulse, followed by some
amplitude oscillations in the center of the pulse, and lastly another
peak symmetric to the first, thus following the red laser pulse, with some
overlap.
For both pulses, the complex phase, shown in the center panel, is close to zero
when there is significant pulse amplitude. In the spectrum (bottom panel), the
overall narrowing and broadening of the red and blue pulse, respectively, is
reflected in a broadening and narrowing of the central peak in the spectrum. The
amplitude modulations on the blue pulse appear as side-lobes in the spectrum.

The initial and final peak of the blue pulse, together with the red pulse are
reminiscent of the counter-intuitive pulse scheme of STIRAP, with the blue laser
acting as the ``Stokes'' pulse and the red laser as ``pump''.
\begin{figure}[tbp] 
  \centering
  \includegraphics{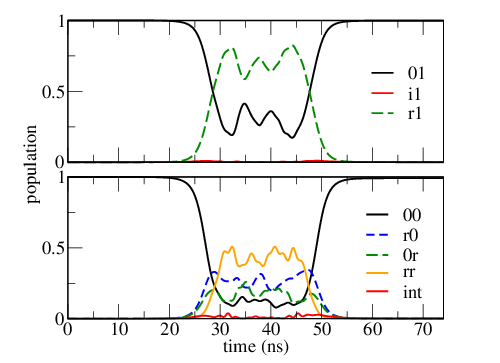}
  \caption{Dissipative population dynamics under the pulse shown in
           Fig.~\ref{fig:pulse:Rydberg:diss}, for the initial states
           $\Op\rho(0) = \Ket{01}\!\Bra{01}$ (top) and
           $\Op\rho(0) = \Ket{00}\!\Bra{00}$ (bottom).
           The intermediary population
           (''int'') is integrated over all levels with  decay, i.e.\
           $\Ket{0i}$, $\Ket{i0}$, $\Ket{ii}$, $\Ket{ir}$, and $\Ket{ri}$.
  }
  \label{fig:dyn:Rydberg:diss}
\end{figure}
The STIRAP-like behavior appears also in the population dynamics, shown
in Fig.~\ref{fig:dyn:Rydberg:diss}, as a population inversion between level
$\Ket{0}$ and $\Ket{r}$, \emph{without} any population in the intermediate
decaying state. The amplitude modulations in the central region of both pulses
then induce some additional dynamics, generating the entanglement needed for the
gate.
Note that the pulse duration for the dissipative process ($T = 75\,$ns) is longer
than that of the coherent process ($T = 50\,$ns). This is necessary to allow for
an adiabatic time evolution that is essential to the STIRAP-like behavior.
Overall, the decaying intermediate state population (red lines in
Fig.~\ref{fig:dyn:Rydberg:diss}) is almost completely suppressed,
which is in contrast to the optimization not taking into account
the dissipation, cf.\ the red lines in Fig.~\ref{fig:dyn:Rydberg:nodiss}.
Both Figs.~\ref{fig:dyn:Rydberg:nodiss} and~\ref{fig:dyn:Rydberg:diss}
show a significant population of the  $\Ket{rr}$ state. This is not
surprising, since the parameters of Table~\ref{tab:atom} are
not in the regime of the Rydberg blockade~\cite{JakschPRL00,SaffmanRMP10}.

\section{Example II: Non-diagonal gates}
\label{sec:allgates}

\begin{table}[tb]
  \centering
 \begin{tabular}{|l|r|}\hline
  qubit frequency  $\omega_1$          & 4.3796~GHz \\ 
  qubit frequency  $\omega_2$          & 4.6137~GHz \\ 
  drive frequency  $\omega_d$          & 4.4985~GHz \\ \hline
  anharmonicity    $\delta_1$          & -239.3~MHz \\ 
  anharmonicity    $\delta_2$          & -242.8~MHz \\ \hline
  effective qubit-qubit coupling $J$   &   -2.3~MHz \\ \hline
  qubit 1 decay time $T_1$             &   38.0~\micro{}s \\
  qubit 2 decay time $T_1$             &   32.0~\micro{}s \\ \hline
  qubit 1 dephasing time $T^{*}_2$     &   29.5~\micro{}s \\
  qubit 2 dephasing time $T^{*}_2$     &   16.0~\micro{}s \\
  \hline
 \end{tabular}
  \caption{Parameters of the transmon Hamiltonian,
    Eq.~\eqref{eq:H_transmon}, and Liouvillian, Eq.~\eqref{eq:L_transmon},
    taken from Ref.~\cite{PolettoPRL12}.}
  \label{tab:transmon}
\end{table}
Superconducting qubits represent a physical realization of a quantum
processor where the Hamiltonian admits both diagonal and non-diagonal
entangling gates. In fact, there exist superconducting
architectures that admit several two-qubit
gates simultaneously~\cite{ChowPRL11,PolettoPRL12}.
We consider here the example of two transmon
qubits coupled via a shared transmission line resonator. In the dispersive
limit, the interaction of each qubit with the resonator leads to an effective
coupling $J$ between the two qubits, and the cavity can be integrated
out~\cite{PolettoPRL12}. The resulting Hamiltonian reads
\begin{eqnarray}
  \label{eq:H_transmon}
  \Op H
  &=& \left(\omega_{1} - \frac{\delta_1}{2} \right) \Op{b}_1^{\dagger} \Op{b}_1
  + \frac{\delta_1}{2} \left( \Op{b}_1^{\dagger} \Op{b}_1 \right)^2
  + \left(\omega_{2} - \frac{\delta_2}{2} \right) \Op{b}_2^{\dagger} \Op{b}_2
  + \frac{\delta_2}{2} \left( \Op{b}_2^{\dagger} \Op{b}_2 \right)^2
  + J \left(\Op{b}_1^{\dagger} \Op{b}_2 +  \Op{b}_1 \Op{b}_2^{\dagger}
  \right)
  \\ &&
  + \Omega(t) \cos \left( \omega_d t \right)
     \left( \Op{b}_1 + \Op{b}_1^{\dagger}
          + \Op{b}_2 + \Op{b}_2^{\dagger}
     \right)\,, \nonumber
\end{eqnarray}
where $\Op{b}_{1,2}$, $\Op{b}_{1,2}^{\dagger}$ are the ladder operators for the
first and second qubit, $\omega_{1,2}$ and $\delta_{1,2}$ represent
the frequency and anharmonicity, $J$ is the effective
qubit-qubit-interaction, and $\Omega(t)$ and $\omega_d$ are amplitude
and frequency of the drive, respectively.
The two most relevant dissipation channels are energy relaxation and pure
dephasing of the qubits, described by the decay rate $\gamma = 1 / T_1$ and
dephasing rate $\gamma_{\phi} = 1 / T_2^{*}$ for each qubit. The
corresponding dissipator reads
\begin{equation}
  \label{eq:L_transmon}
  \mathcal L_D(\Op\rho)
  = \sum_{q=1,2} \left(
    \gamma_q \sum_{i=1}^{N-1} i D\left[\Ket{i-1}\!\Bra{i}_q\right] \Op\rho
    +
      \gamma_{\phi, q} \sum_{i=0}^{N-1}
      \sqrt{i} D\left[\Ket{i}\!\Bra{i}_q\right]
      \Op\rho
    \right)\,,
\end{equation}
with
\begin{equation}
  D\left[\Op{A} \right] \Op{\rho}
  = \Op{A} \Op{\rho} \Op{A}^{\dagger}
    - \frac{1}{2} \left(
      \Op{A}^{\dagger} \Op{A} \Op{\rho}
      + \Op{\rho} \Op{A}^{\dagger} \Op{A}
    \right)
\end{equation}
and each qubit, $q = 1,2$, truncated at level $N$.
The parameters of the coupled transmon qubits are summarized in
Table~\ref{tab:transmon}. We employ a
RWA, centered at the drive frequency $\omega_d$.

\begin{figure}[tb] 
  \centering
 \includegraphics{tm_full_diss_f_avg}
 \caption{Optimizing a \sqrtISWAP{} gate for two transmons in the
   presence of energy relaxation and pure dephasing (with the rates
   given in Table~\ref{tab:transmon}): Convergence
   for five choices of sets of initial states, as described in the text. The gate
   duration is $T = 400\,$ns. The panels from top to bottom show the
   gate error over the number of iterations; the
   gate error over the number of state propagations,
   indicative of the required CPU time; a zoom on the
   initial phase of the optimization; and a zoom on the asymptotic
   convergence (panels c and d both using a linear scale). The number
   of propagations (x-axis in panels b-d) is a linear rescaling of the number of
   OCT iterations (x-axis in panel a), with 2 propagations per iteration and
   state, i.e., the lines of panel a are rescaled differently depending
   on the respective number of states. Since all panels only show different
   views on the same data, the line colors and styles are the same in all of
   them.}
 \label{fig:converg:transmon_full_diss}
\end{figure}
The Hamiltonian in Eq.~(\ref{eq:H_transmon}) can generate a large number of
entangling two-qubit gates; we find \sqrtISWAP{} to be a fast converging
non-diagonal perfect entangler, and thus choose
\begin{equation}
  \Op{O}
  = \begin{pmatrix}
    1 &                  0 &                  0 & 0 \\
    0 & \frac{1}{\sqrt{2}} & \frac{i}{\sqrt{2}} & 0 \\
    0 & \frac{i}{\sqrt{2}} & \frac{1}{\sqrt{2}} & 0 \\
    0 &                  0 &                  0 & 1
  \end{pmatrix}
  \label{eq:sqrt_iswap}
\end{equation}
as the optimization target.
Figure~\ref{fig:converg:transmon_full_diss} shows the convergence behavior for
several choices of initial states: the 16 canonical states of the full
basis of Liouville space; the three states given in
Eq.~(\ref{eq:rhos}) with equal weight and
with $w_1 / w_2 = w_1 / w_3 = 20$; a set of 5 states consisting of $\Op\rho_1$
expanded into four pure states, cf. Eq.~\eqref{eq:rho1_dp1}
plus $\Op\rho_2$ of Eq.~\eqref{eq:rho2}; and lastly a set of eight
states, cf. Eqs.~\eqref{eq:rho1_dp1}
and~\eqref{eq:rho2_2d},
consisting of the expansion of $\Op\rho_1$ and the four pure
states of a mutually unbiased basis, as explained in
Section~\ref{sec:oct}.
As seen in the top panel, all choices 
show good convergence. A plateau corresponding to a slowing of
convergence is observed only for the three
states with equal weights. But even in this case, the same asymptotic
value for the gate error is obtained as for the other choices, see also
Fig.~\ref{fig:converg:transmon_full_diss}(d).
The advantage of employing the reduced sets of
states in the optimization functional, Eq.~\eqref{eq:functional},
becomes most apparent in Fig.~\ref{fig:converg:transmon_full_diss}(b)
which shows the gate error over the number of state
propagations. Since optimization requires two propagations per
iteration and state, i.e., the
backward and forward propagation in Eq.~\eqref{eq:control}, the number
of state propagations corresponds directly to the CPU time that is
required to obtain a given
fidelity. Figure~\ref{fig:converg:transmon_full_diss}(c) and (d)
shows a zoom on the same data, once for the initial phase
of the optimization and once for the asymptotic behavior. All reduced sets
except for the three states with equal weights perform better than the
full set during the
initial phase.  Also, for this specific optimization problem, all reduced sets
reach a slightly better asymptotic value than the full set, although we
expect that ultimately all curves will converge to the same value.
Figure~\ref{fig:converg:transmon_full_diss} suggests that
the reduced sets have a significant
advantage in reaching a good fidelity with a given amount of resources, especially
since in practice, an optimization is usually stopped near the beginning of the
asymptotic regime. Indeed, the full set shows an advantage only in the
intermediate regime between gate errors of 10 and 1 percent, and only over the
sets of three states. The choice of 5 or 8 states outperforms the full set in
all cases.
One should note that the savings in computational resources due to the
use of a reduced set of
states also extend to the amount of memory required, which is proportional to
the number of states. Since in the optimization algorithm, propagated states
over the entire time grid need to be stored, these savings can be very
substantial.

For the three states with equal weights the gate error
shows a non-monotonic behavior in the upper left corner of
Fig.~\ref{fig:converg:transmon_full_diss}(c). This is due to the
optimization functional, Eq.~\eqref{eq:functional}, not being
equivalent to the gate error $F_{\avg}$,
Eq.~\eqref{eq:Favg}. Specifically, for a set of three states, no bound
on the distance between $J_T$  and $1-F_{\avg}$ can be
derived~\cite{ReichKochPRA13}. Thus,
the gate error might increase even though $J_T$
decreases. In fact, the behavior of $J_T$ is fully monotonic as
expected (data not shown).  With an increasing number of states in the chosen set, the
value of the optimization functional is more closely connected to the
gate fidelity; and for 5 and 8 states numerical, respectively
analytical, bounds can be found~\cite{ReichKochPRA13, HofmannPRL05}.
For this reason, we expect the sets of 5 and 8 states to show
a faster convergence than the 3 states, when measured in OCT
iterations, although not
necessarily in CPU time. This expectation is confirmed by
Fig.~\ref{fig:converg:transmon_full_diss}.
The weak correspondence between the optimization functional and
the gate error for three states is most likely also the reason for the
plateau observed for the red dashed line in
Fig.~\ref{fig:converg:transmon_full_diss}(a) and (b).
However, the use of three states can still be a good choice since
weighting the states properly improves the
convergence significantly. The weights have to be chosen empirically,
but the choice can be guided by physical intuition.  The
three states are responsible for ensuring that the realized gate is diagonal in
the correct basis, that the relative phases match the target once the correct
basis has been found, and that the gate is unitary on the logical subspace,
respectively. The weights should reflect which of these requirements
is most difficult to realize.
In the present example this is finding the correct basis in which the gate is
diagonal. Therefore the choice of $w_1/w_2 = w_1/w_3 = 20$ gave the best
convergence rate. This is in contrast to the optimization of the Rydberg gate in
Section~\ref{sec:phasegate}, in which the gate was already known to be diagonal,
and the first state could be left out of the optimization entirely. Generally,
using the set of three states with equal weights is not recommended.

Comparing Fig.~\ref{fig:converg:transmon_full_diss} with the bottom panel of
Fig.~\ref{fig:converg:Rydberg} for the Rydberg gate shows that the different
choices of basis sets show a slightly wider range of the convergence rate.
This can be attributed to the fact that for the Rydberg gate, the
optimization landscape is severely constrained since only
diagonal gates can be reached. In contrast, the transmon Hamiltonian
can generate both diagonal
and non-diagonal gates, resulting in a more complex optimization landscape.
Different choices of initial states can thus take more strongly varying pathways.

\begin{figure}[tb] 
  \centering
 \includegraphics{tm_weak_diss_f_avg}
 \caption{Optimizing a \sqrtISWAP{} gate for two transmons with weak
   dissipation, using
   decay and dephasing times increased by a factor of 10 compared to
   Fig.~\ref{fig:converg:transmon_full_diss} (with all quantities and
   labels as defined in Fig.~\ref{fig:converg:transmon_full_diss}). The gate duration
   is $T=400\,$ns.  The weaker dissipation results in an asymptotic gate error of
   approximately $7\cdot10^{-4}$ compared to $7\cdot10^{-3}$ in
   Fig.~\ref{fig:converg:transmon_full_diss}, cf.\ the y-axis scales in
   both figures.
   }
 \label{fig:converg:transmon_weak_diss}
\end{figure}
Figure~\ref{fig:converg:transmon_weak_diss} shows the optimization of
a \sqrtISWAP{} gate for two transmons in the
case of weak dissipation, where the decay and dephasing times from
Table~\ref{tab:transmon} have been increased by a factor of 10. A comparison of
Fig.~\ref{fig:converg:transmon_weak_diss}(a) with
Fig.~\ref{fig:converg:transmon_full_diss}(a) shows that the
convergence
behavior is essentially the same except for the value of the asymptote. We find
an asymptotic gate error of approximately $7 \cdot 10^{-3}$ with full
dissipation, $7 \cdot 10^{-4}$ with weak dissipation, and no asymptote without
dissipation (data not shown). The value of the asymptote is logarithmically
proportional to the decay and dephasing rates. This is as expected
since the pulse duration is kept constant at $400$~ns and the gate
fidelity is solely limited by
dissipation. Our claim that the dissipation only affects the asymptotic
convergence is supported by a comparison of the initial convergence
in Figs.~\ref{fig:converg:transmon_full_diss}(c)
and~\ref{fig:converg:transmon_weak_diss}(c),
which remarkably are completely identical.
Furthermore, the crossing between the black solid and red dot-dashed lines for
the full basis and the three states with unequal weights near 1000 propagations
and that between the blue dotted and orange dash-dash-dotted lines for the sets
of 5, respectively 8, states near
1300 propagations in Fig.~\ref{fig:converg:transmon_weak_diss}(d) can
also be seen in Fig.~\ref{fig:converg:transmon_full_diss}(d).
There are however some slight differences in the asymptotically reached values,
in that the choice of 3 states (with both equal and unequal weights) reaches a slightly
smaller gate error than in the case of full dissipation. Again, we expect that
ultimately, all curves will converge to the same value. Which set of states
reaches the best gate error at a specific point near the beginning of the
asymptotic region seems to depend on the slope of the convergence curve as the
limit is approached. This can depend on any number of factors
including, e.g., the choice
of $\lambda_a$ in Eq.~(\ref{eq:J}). Again, empirically, the reduced sets of
states show a significant numerical advantage over the full basis also
for weak dissipation.

\begin{figure}[tbp] 
  \centering
  \includegraphics{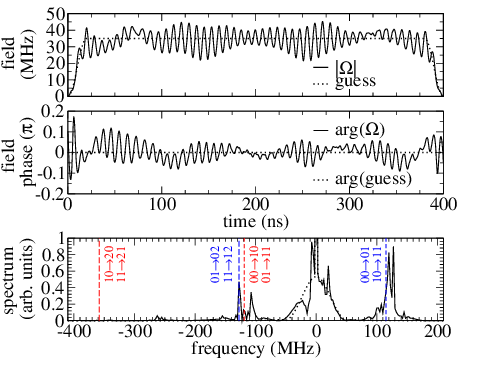}
  \caption{Shape and spectrum of an optimized pulse, from optimization with
    3 weighted states, with strong dissipation. The panels from top to bottom show the
    amplitude, complex phase, and spectrum of the optimized
    pulse $\Omega(t)$.  The spectrum is shown in the rotating frame, with
    zero corresponding to the driving frequency $w_d$ of the field. The
    transition frequencies from the logical subspace are indicated by
    vertical dashed lines.
    These are $\Delta_1 = w_1 - w_d = -118.88\,$MHz and
    $\Delta_1 - \delta_1 = -358.18\,$MHz in red for the left qubit, and
    $\Delta_2 = w_2 - w_d = 115.20\,$MHz and
    $\Delta_2 - \delta_2 = -127.58\,$MHz in blue for the right qubit.
    The central peak in the spectrum has been cut off to show the
    relevant side-peaks, and would extend to a value of approximately
    10.0.  For all quantities, the values for the guess pulse are shown
    as a dotted line.
  }
  \label{fig:transmon_pulse}
\end{figure}
\begin{figure}[tb] 
  \centering
  \includegraphics{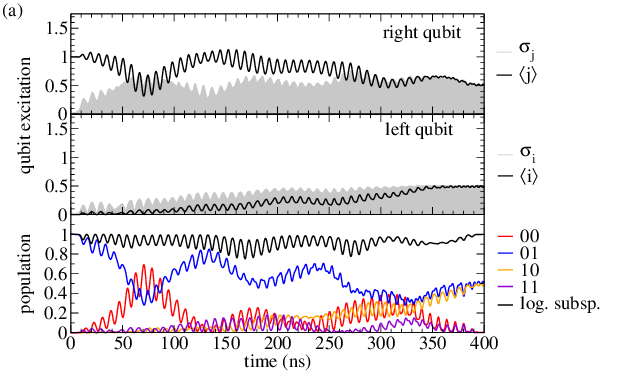}
  \includegraphics{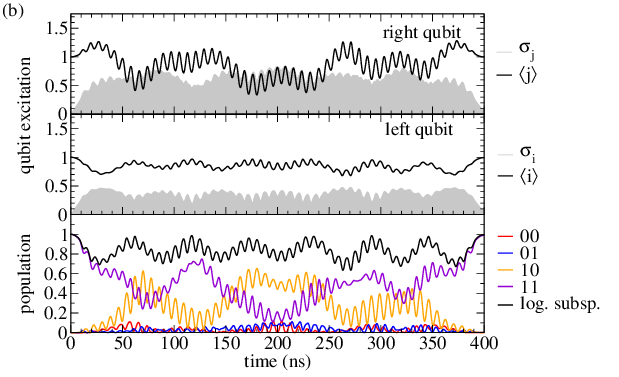}
  \caption{Population dynamics for
          $\Op\rho(t=0) = \Ket{01}\!\Bra{01}$ (a) and
          $\Op\rho(t=0) = \Ket{11}\!\Bra{11}$ (b) under the
          pulse shown in Fig.~\ref{fig:transmon_pulse}. For each of the two
          propagated states, the expectation value of the right qubit excitation
          quantum number $j$ is shown in the top panel, with the standard
          deviation in gray, the expectation value for the corresponding quantum
          number $i$ for the left qubit is shown in the center panel, and the
          population dynamics for all the logical subspace states is shown in
          the bottom panel (colored lines), along with the total population in
          the logical subspace (black line).}
  \label{fig:transmon_popdyn}
\end{figure}
As an example, the optimized pulse obtained using a set of  three
states with unequal weights, taking into account the full dissipation,
is presented in
Fig.~\ref{fig:transmon_pulse}, along with the pulse spectrum. The population
dynamics that this pulse induces when propagating the logical basis states
$\Op{\rho}(t=0) = \Ket{01}\!\Bra{01}$ and  $\Op{\rho}(t=0) = \Ket{11}\!\Bra{11}$
is shown in Fig.~\ref{fig:transmon_popdyn}.
As can be seen in the top panel of Fig.~\ref{fig:transmon_pulse}, the optimized
pulse shows small oscillations around the guess peak amplitude of 35~MHz. The complex
phase, shown in the middle panel, stays relatively close to zero, indicating
that the optimization employs mainly amplitude modulation. The pulse
amplitude is roughly time-symmetric. The pulse spectrum shown in the
bottom panel of Fig.~\ref{fig:transmon_pulse}
relates easily to the pulse shape. The strongest frequency component remains the
driving frequency of the guess pulse (zero in the spectrum). The small
oscillations in the pulse shape are approximately 8$\,$ns apart, corresponding to
a frequency of $\pm$125$\,$MHz, which is present in the spectrum. There are peaks
with exponentially decaying amplitude in the spectrum at multiples of these
values. The width of the central peak is due to the 20$\,$ns switch-on and
switch-off time of the pulse, and is unchanged from the guess pulse. The
fact that there is not a single, but a double peak around $\pm$125$\,$MHz
corresponds the slow beats in the pulse shape.  The slight asymmetry of the
spectrum is caused by the complex phase of the optimized pulse.

The spectrum of the optimized pulse is very instructive in understanding the
population dynamics in Fig.~\ref{fig:transmon_popdyn}. The most relevant
transition frequencies from the logical subspace are indicated by vertical lines
in the spectrum in the lower panel of Fig.~\ref{fig:transmon_pulse}. Clearly,
the peaks around $\pm$125~MHz are nearly resonant with the excitation of the
left and right qubit, and the excitation to level $\Ket{2}$ of the right qubit.
There is no significant component in the spectrum that could excite to the level
$\Ket{2}$ of the left qubit.
Consequently, in the population dynamics of both the $\Ket{01}\!\Bra{01}$ and
$\Ket{11}\!\Bra{11}$ state, the right qubit (top panel) leaves the logical
subspace (expectation value $\langle j\rangle>1.0$) to a much more
significant extent than the
left qubit (middle panel). This behavior is slightly more pronounced for
$\Ket{11}\!\Bra{11}$, which is the only state for which the total subspace
population (gray curve in bottom panel) drops below 80\% for a significant
amount of time.
The fact that for all logical basis states, most of the dynamics
occurs within the logical
subspace is due to the presence of decoherence, where higher levels have
faster decay and faster dephasing due to a stronger coupling to the cavity. In an
optimization without dissipation (data not shown), the optimized dynamics would
generally veer farther outside the logical subspace. Lastly, the population
dynamics show the expected behavior for the \sqrtISWAP{} gate: the $\Ket{01}$
state ends up in a coherent superposition between $\Ket{01}$ and $\Ket{10}$,
whereas $\Ket{11}$ returns to its original state at the end of the gate.

\section{Conclusions}
\label{sec:concl}

We have utilized the fact that the average error of a quantum gate can
be estimated from the time evolution of a reduced set of
states~\cite{ReichKochPRA13,ReichKochPRL13} to
construct a dedicated functional for quantum gate optimization in
open quantum systems. Our
optimization functional consists of Hilbert-Schmidt products that
compare the actual and ideal time-evolved states from the reduced set. The
minimal number of states that need to be forward and backward
propagated during optimization is two for Hamiltonians that admit only
diagonal gates and three for Hamiltonians that allow for both diagonal
and non-diagonal gates. Remarkably, the size of the minimal set of
states is independent of Hilbert space dimension.

While the minimal number of states allows for determining whether
a quantum gate has been implemented, it is insufficient to deduce
bounds on the gate error~\cite{ReichKochPRA13}. Numerical bounds
require $d+1$ states in the reduced set, where $d$ is the dimension of
the Hilbert space on which the optimization target is defined. In
order to obtain meaningful analytical bounds on the gate error, $2d$ states are
necessary. Employing the
sets of $d+1$, respectively $2d$, states in quantum gate optimization
is still significantly
more efficient, both with respect to CPU time and memory requirements,
than utilizing a full basis of Liouville space, with
$d^2$ elements~\cite{KallushPRA06,OhtsukiNJP10,ToSHJPB11}.

We have demonstrated the power of our approach in the optimization of
a diagonal and a non-diagonal two-qubit gate. Specifically, we have
optimized a controlled phasegate for trapped neutral atoms that are
excited into a Rydberg state and subject to fast spontaneous emission
from an intermediate state. The best performance was achieved by two
states in the reduced set and a large weight of the Hilbert-Schmidt
product for the state responsible for detecting
phase errors. In the optimization of a \sqrtISWAP{} gate for two
transmons coupled to the same transmission line cavity and subject to
both energy relaxation and pure dephasing, we have
found the best, and roughly identical, performance for the reduced
sets consisting of $d+1$, respectively $2d$, states. In all cases, the
final gate error was limited by the decoherence rates. This confirms
that employing a reduced set of states in quantum gate optimization is
sufficient to determine the physical limit for the gate
error.

The significant reduction in computational resources that we report
here opens the door for a large-scale, systematic investigation of the
fundamental limits of high-fidelity quantum gates in the presence of
decoherence. Our approach is not tied to a specific decoherence
model. It therefore allows to explore, using optimal control theory,
settings for extended Hilbert spaces and beyond Markovian master
equations, where a quantum system's complexity
may possibly be exploited for control.

\begin{acknowledgments}
  We would like to thank Alberto Castro,
  Giulia Gualdi,
  Matthias~M.~Müller,
  Felix Motzoi, Alireza Shabani and Birgitta
  Whaley for fruitful discussions and the Kavli Institute
  of Theoretical Physics at the University of California at Santa
  Barbara for hospitality.
  This research was supported in part by the Deutscher Akademischer
  Austauschdienst and by the National Science Foundation
  (Grant No. NSF PHY11-25915).
\end{acknowledgments}

\appendix
\pagebreak

\section{Three states are sufficient to assess whether a desired target unitary is implemented}
\label{sec:proof}

In the following we discuss the functional $J_{dist}$,
\begin{equation}
  J_{dist} = \sum_{i=1}^{3}
   \Tr\left[\left(\Op O\Op\rho_{i}(0)\Op
       O^{\dagger}-\Op\rho_{i}\left(T\right)\right)^2\right]\,,
  \label{eq:distance_functional}
\end{equation}
which is built on the distance between the ideal and actual
states at time $T$. It attains its global minimum, $J_{dist}=0$,
if and only if the initial states defined in Section~\ref{sec:oct},
$\Op\rho_i(0)$ for $i=1,2,3$, are mapped to their correct target
states, i.e., fulfill condition~\eqref{eq:cond}.
This functional motivates the use of the optimization functional
$J_T$, Eq.~\eqref{eq:functional},
which is also built on only three states, as discussed in
Sec.~\ref{subsec:funct}. $J_T$ and $J_{dist}$ differ in that
$J_T$ evaluates the Hilbert-Schmidt
products, i.e., the projections of the actual onto the ideal states
instead of the trace distance.
The construction of $J_{dist}$, and subsequently $J_T$,
is rationalized by a theorem for unital, i.e., identity preserving, dynamical
maps. Specifically, the theorem states that a \emph{complete and totally
rotating} set of density matrices is sufficient to determine whether
a given time evolution is unitary. The
functional~\eqref{eq:distance_functional} exploits
the further property of a complete and totally rotating set
of density matrices to differentiate any two
unitaries~\cite{ReichKochPRA13}. The theorem for unital dynamical maps
is proven in Sec.~\ref{subsec:proofs}.

It should be stressed that we use $J_T$, Eq.~\eqref{eq:functional},
instead of $J_{dist}$, Eq.~\eqref{eq:distance_functional},
as optimization functional. This is motivated by the convexity of
$J_T$ which implies a much more favorable convergence behavior than
would be obtained with a non-convex functional~\footnote{Optimization
  using non-convex functionals is possible but requires additional
  terms in the update equation for the field to preserve monotonicity
  of the convergence~\cite{ReichKochJCP12}.
}. Mathematically, however, the two functionals are not equivalent.
This is illustrated by rewriting a single summand of $J_{dist}$,
Eq.~\eqref{eq:distance_functional}, and comparing it to
the corresponding term in $J_T$, Eq.~\eqref{eq:functional},
\begin{equation}
  \Tr\left[\left(\Op O\Op\rho_{i}(0)\Op
      O^{\dagger}-\Op\rho_{i}\left(T\right)\right)^2\right]
  = \Tr\left[\left(\Op O\Op\rho_{i}(0)\Op
      O^{\dagger}\right)^2\right]-2\Tr\left[\Op O\Op\rho_{i}(0)\Op
    O^{\dagger}\Op\rho_{i}\left(T\right)\right]
   + \Tr\left[\left(\Op\rho_{i}\left(T\right)\right)^2\right]\,.
  \label{eq:functional_expansion}
\end{equation}
The first term on the rhs of Eq.~\eqref{eq:functional_expansion} is
constant and thus irrelevant. The second term corresponds
to the Hilbert-Schmidt overlap as used in $J_T$,
Eq.~\eqref{eq:functional}, up to a prefactor. The main difference between
$J_T$ and $J_{dist}$ is due to the third term,
the purity of the propagated density matrix.
$J_T$ neglects this term. This could potentially disturb convergence,
because  the functional value of $J_T$ can be decreased
by (artificial) purification of the totally mixed states $\Op\rho_1$ and
$\Op\rho_3$, cf. Eq.~\eqref{eq:rhos}, instead of being decreased due
to the desired approach to the target.
Note that this problem can only arise for mixed states, i.e., when
using the minimal set of states. For the reduced sets consisting of $d+1$,
respectively $2d$, states, propagation starts from pure states, and the global
minimum of $J_T$ is identical to the global minimum of $J_{dist}$.
Note that the problem of artificial purification is purely
hypothetical and was never encountered
in our optimizations --  'artificial purification traps' in
the optimization landscape of the functional
$J_T$ with mixed states are apparantly avoided.

\subsection{Construction of the functional}
\label{subsec:funct}

We first define the concept of \emph{complete and total
rotation}, which we then use to formulate the required theorem.
Let $\mathcal{H}$ be a Hilbert space with dimension $N$. Let
$\mathcal{A}$ be a set of
$N$ one-dimensional orthogonal projectors. A one-dimensional projector
is a projector with rank one, which means that its spectrum consists of a
single eigenvalue equal to one with all remaining eigenvalues being zero.

\textbf{Definition}:
A one-dimensional projector $\Op P_{TR}$ is called \emph{totally rotated} with
respect to the set $\mathcal{A}$ if $\forall \Op P\in\mathcal{A}:\
\Op P_{TR}\Op P\neq0$.

\textbf{Definition}:
A set of density operators, $\left\{ \Op\rho_{i}\right\}$ with
$\Op\rho_{i}\in\mathcal{H}\otimes\mathcal{H}$, is called \emph{complete} if
the set $\mathcal{P}$ of projectors onto the eigenspaces of
$\left\{ \Op\rho_{i}\right\}$ contains exactly
$N$ one-dimensional orthogonal projectors.

\textbf{Definition}:
A set of density operators, $\left\{ \Op\rho_{i}\right\}$ with
$\Op\rho_{i}\in\mathcal{H}\otimes\mathcal{H}$, is called \emph{complete and
totally rotating} if it is complete and there exists a
one-dimensional projector in $\mathcal{P}$ that is totally rotated
with respect to the orthogonal set of one-dimensional orthogonal
projectors necessary for completeness.

\textbf{Theorem 1}:
Let $\text{DM\ensuremath{\left(N\right)}}$ be the space of $N\times N$
density matrices and $\mathcal{D}:\text{DM\ensuremath{\left(N\right)}}
\mapsto\text{DM\ensuremath{\left(N\right)}}$ a
dynamical map. The following three statements are equivalent:
\begin{enumerate}
\item $\mathcal{D}$ is unitary, i.e.,
  $\mathcal{D}\left(\rho\right)=U\rho U^{\dagger}$
  $\forall\rho\in\text{DM\ensuremath{\left(N\right)}}$
  and $U$ some element of the projective unitary group,
  $U\in\text{PU\ensuremath{\left(N\right)}}$.
\item $\mathcal{D}$ maps a set $\mathcal{A}$ of $N$ one-dimensional
  orthogonal projectors onto a set of $N$ one-dimensional orthogonal
  projectors as well as a totally rotated projector $\Op P_{TR}$
  (with respect to $\mathcal{A}$) onto a one-dimensional projector.
\item $\mathcal{D}$ is unital and leaves the spectrum of a complete
  and totally rotating set of density matrices invariant.
\end{enumerate}

We now explain how Theorem~1 can be used to prove the claim that
$J_{dist}$, Eq.~\eqref{eq:distance_functional},
attains its global minimum if and only if condition~\eqref{eq:cond}
is fulfilled for the three states defined in Sec.~\ref{sec:oct}. We
first discuss the role of $\Op\rho_3 = \frac{1}{N} \openone$. It is used
to check whether the evolution corresponds to a dynamical map in the
optimization subspace and whether it is unital. This
dynamical map is obtained by
projecting the action of the dynamical map, defined on the total
Hilbert space,
onto the optimization subspace.
The term in the functional~\eqref{eq:distance_functional} involving $\Op\rho_3 =
\frac{1}{N} \openone$ becomes minimal, and so does the total
functional, only if the identity in the optimization subspace is
mapped onto itself. Minimization of $J_{dist}$ thus ensures a unital
dynamical map on the subsystem such that Theorem~1 is applicable.

We now discuss the role of $\Op\rho_1$ and $\Op\rho_2$ which by construction
form a complete and totally rotating set of density matrices. The
functional~\eqref{eq:distance_functional}
becomes zero only if $\mathcal{D}(\Op\rho_1) = \Op O \Op\rho_1 \Op
O^\dagger$ and $\mathcal{D}(\Op\rho_2) = \Op O \Op\rho_2 \Op
O^\dagger$. This requires the actual evolution to be unitary. Unitary
evolution leaves the spectrum of a density matrix invariant. Due to
the equivalence relation $(1)\Longleftrightarrow (3)$ in Theorem~1,
preservation of
the spectrum of a complete and totally rotating set of density matrices, i.e.,
the two states $\Op\rho_1$ and $\Op\rho_2$, is sufficient to ensure
unitarity.
Furthermore, it was proven in Ref.~\cite{ReichKochPRA13} that the density
matrices $\Op\rho_{1}$ and $\Op\rho_{2}$ are
unitary differentiating, i.e., it is possible to distinguish any two
unitary evolutions by inspection of $\Op\rho_{1}(T)$ and $\Op\rho_{2}(T)$
only. In particular there is only one unitary dynamical map,
$\mathcal{D}(\Op\rho) = \Op U \Op\rho \Op U^\dagger$, which leads to
$\mathcal{D}(\Op\rho_{i}(0)) = \Op O \Op\rho_i \Op O^\dagger$ for both $
i=1,2$, namely the one induced by the target unitary $\Op O$. Therefore
the functional~\eqref{eq:distance_functional} becomes minimal if and only if
the target gate $\Op O$ is implemented.

To summarize, $J_{dist}$ is additively composed of three terms,
each corresponding to a distance measure between the desired result,
$\Op O \Op\rho_i \Op O$, and the actually implemented evolution,
$\mathcal{D}(\Op\rho_i)$. For the total functional to be minimal, the
evolutions of all three states have to match. This
is the case only if a unital dynamical map on the optimization
subspace is implemented \emph{and} if this is the unitary evolution
according to $\Op O$. More explicitly, the distance measure formed by the
density matrices $i=1,2$ is only meaningful provided the evolution
within the optimization subspace corresponds to a unital dynamical map.
However, this is ensured by the third density matrix. Consequently, the
global minimum of the functional~\eqref{eq:distance_functional} will
only be attained if this condition is fulfilled, too.

Note that the functional~\eqref{eq:distance_functional} weights
all three states equally. This is not a unique choice. In fact, all
crucial properties of the functional remain unchanged when scaling the
three terms with different positive factors, which has been done in the
main text for example when discussing the optimisation using three states
with weighting which significantly improved the performance of the
optimization.

\subsection{Proof}
\label{subsec:proofs}

We utilize in the following the representation of operators by
$N\times N$ matrices and therefore omit the operator notation.
In order to prove Theorem 1, it is useful
to first show the validity of the following lemma.

\textbf{Lemma 1}:
Let $\mathcal{D}$ be a unital dynamical map, i.e.,
$\mathcal D$ is completely positive and maps identity onto itself,
acting on $N \times N$
density matrices. If and only if there exists a set of $N$
one-dimensional, orthogonal projectors that is mapped by $\mathcal{D}$
onto another set of $N$
one-dimensional orthogonal projectors,
there exists a complete set of density matrices whose spectrum is invariant
under $\mathcal{D}$.

\textbf{Proof of Lemma 1}: ($\Longrightarrow$ direction) We denote the set of
$N$ one-dimensional projectors $P_{i}$ by $\mathcal{P}$.
By assumption, we know that
\begin{equation*}
\forall i:\ \mathcal{D}\left(P_{i}\right)=\tilde{P}_{i} \,,
\end{equation*}
where the $\tilde{P}_{i}$ also form a set of $N$
one-dimensional, orthogonal projectors.
Clearly, $\text{spec}\left(P_{i}\right)=\text{spec}(\tilde{P}_{i})$,
hence $\forall P_i\in\mathcal{P}$
\begin{equation*}
\text{spec}\left(\mathcal{D}\left(P_i\right)\right)=
\text{spec}\left(P_i\right) = (1,0,\dots,0)\,.
\end{equation*}
Obviously, $\mathcal{P}$ itself corresponds to a specific complete set of
density matrices, $\rho_i=P_i$.

($\Longleftarrow$ direction)
This part of the proof proceeds as follows: First we show that the
assumption, a dynamical map leaving the spectrum of a given density
matrix invariant, implies that $\mathcal D$ maps projectors onto
the eigenspaces of the initial density matrices
into projectors onto the eigenspaces of the resulting density matrix
with the same eigenvalue. As a consequence, a one-dimensional projector
onto a corresponding one-dimensional eigenspace is mapped into a
one-dimensional projector.
We then repeat this argument for all density matrices in the complete
set. In this set, by definition, there exist density matrices with $N$
one-dimensional, orthogonal projectors onto one-dimensional
eigenspaces which, according to the first step of the $\Longleftarrow$
proof, is mapped onto another set of one-dimensional projectors. We
show in a second step
that the set of the mapped one-dimensional projectors is also orthogonal.

We start by assuming that $\mathcal{D}$ leaves the
spectrum of a given density matrix, $\rho$, invariant,
\begin{equation*}
\text{spec}\left(\mathcal{D}\left(\rho\right)\right)=
\text{spec}\sum_{k}\left(E_{k}\rho E_{k}^{\dagger}\right)=
\text{spec}\left(\rho\right) \,,
\end{equation*}
where we have expressed $\mathcal{D}$ in terms of Kraus operators
$E_k$. We can write $\rho=\sum_{i}\lambda_{i}P'_{i}$ where
$\mathcal{P}'=\left\{P'_{i}\right\}$ is a set of $M$ orthogonal
projectors onto the eigenspaces of $\rho$ with $M$ the number of
distinct eigenvalues of $\rho$.
We assume the $\lambda_{i}$ to be ordered by magnitude
with $\lambda_{1}$ corresponding to the largest eigenvalue.
Since we know that the spectrum of $\mathcal{D}\left(\rho\right)$
to be identical to that of $\rho$, we can
decompose $\mathcal{D}\left(\rho\right)$,
\begin{eqnarray*}
\mathcal{D}\left(\rho\right) & = & \sum_{i}\lambda_{i}\tilde{P}'_{i}
\end{eqnarray*}
with $\{ \tilde{P}'_{j}\} $ another set of $M$ orthogonal
projectors.
Note that neither the $P'_{i}$ nor the $\tilde{P}'_{i}$
have to be one-dimensional but
for a given $i$, $\tilde{P}'_{i}$ has the same dimensionality
as the corresponding $P'_{i}$. Specifically,
\begin{equation*}
\mathcal{D}\left(\rho\right)=
\mathcal{D}\left(\sum_{i}\lambda_{i}P'_{i}\right)=
\sum_{i}\lambda_{i}\mathcal{D}\left(P'_{i}\right)=
\sum_{j}\lambda_{j}\tilde{P}'_{j} \,.
\end{equation*}
Multiplying by another projector $\tilde{P}'_{p}$ from the set,
where $p$ can take integer values between $1$ and $M$,
we obtain
\begin{equation}
\sum_{i}\lambda_{i}\mathcal{D}\left(P'_{i}\right)\tilde{P}'_{p}
=\sum_{j}\lambda_{j}\tilde{P}'_{j}\tilde{P}'_{p}
=\lambda_{p}\tilde{P}'_{p} \,, \label{eq:startingequation}
\end{equation}
since $\tilde{P}'_{j}$, $\tilde{P}'_{p}$ are orthogonal.
Using proof by (transfinite) induction we now show that
\begin{equation*}
\mathcal{D}\left(P'_{k}\right)=\tilde{P}'_{k} \quad \forall i=k,\dots, M\,.
\end{equation*}
The idea of the induction is the following: To show that indeed the
projectors onto the eigenspaces of $\rho$, $P'_i$, are mapped into
projectors onto the eigenspaces of $\mathcal{D}(\rho)$ with the same
eigenvalue, we start with the projector onto the eigenspace
with the largest eigenvalue and then inductively proceed to
increasingly smaller eigenvalues. Furthermore, to prevent having to
deal with a possible smallest eigenvalue of $0$, we treat the
lowest eigenvalue case separately.
Calling the induction variable $k$,
we have to show that $\mathcal{D}\left(P'_{k}\right)=\tilde{P}'_{k}$
follows from the assumption
$\mathcal{D}\left(P'_{i}\right)=\tilde{P}'_{i}~\forall
i<k$.
Note that if $k=M$, i.e., for the smallest eigenvalue,
\begin{equation}
\label{eq:unital}
\sum_{i}\mathcal{D}\left(P'_{i}\right)
=\mathcal{D}\left(\sum_i P'_{i}\right)
=\mathcal{D}\left(\openone\right)=\openone\,,
\end{equation}
since, by definition, a unital dynamical map maps identity
onto itself. So assume $k\neq M$. Then $\lambda_{k}>0$ since it
is not yet the smallest eigenvalue because each $\lambda{k}$ corresponds
by construction to a different eigenspace, hence they are different, and
we assumed them to be ordered.
For $k=p$, we can rewrite Eq.~\eqref{eq:startingequation},
multiplying by an arbitrary normalized eigenvector
$\vec{x}_{k}\in\mathbb{C}^N$ of $\tilde{P}'_{k}$ from the left and right,
\begin{equation}
\label{eq:lambdas}
\sum_{i}\lambda_{i}\vec{x}_{k}\cdot\mathcal{D}\left(P'_{i}\right)\cdot\vec{x}_{k}
=\lambda_{k} \,.
\end{equation}
By assumption of the induction,
$\mathcal{D}\left(P'_{i}\right)=\tilde{P}'_{i}~\forall i<k$, therefore
\begin{equation*}
  \vec{x}_{k}\cdot\mathcal{D}\left(P'_{i}\right)\cdot\vec{x}_{k}=0
  \quad \forall i<k\,.
\end{equation*}
Introducing $d_{kk}^{\left(i\right)}
\equiv \vec{x}_{k}\cdot\mathcal{D}\left(P'_{i}\right) \cdot\vec{x}_{k}$,
Eq.~\eqref{eq:lambdas} can be written as
\begin{equation}
\label{eq:secondequation}
\sum_{i\geq k}\lambda_{i}d_{kk}^{\left(i\right)}=\lambda_{k} \,.
\end{equation}
Due to Eq.~\eqref{eq:unital} and the assumption of the induction,
\begin{eqnarray*}
\sum_{i}d_{kk}^{\left(i\right)}
=\sum_{i\geq k}d_{kk}^{\left(i\right)}
=\sum_{i}\vec{x}_{k}\cdot\mathcal{D}\left(P'_{i}\right)\cdot\vec{x}_{k}
= 1\,,
\end{eqnarray*}
and, since $\mathcal{D}\left(P'_{i}\right)$ is the image of a positive
semidefinite matrix which has to be positive semidefinite
itself,
\begin{eqnarray*}
d_{kk}^{\left(i\right)}=\vec{x}_{k}\cdot\mathcal{D}\left(P'_{i}\right)
\cdot\vec{x}_{k} \geq  0\ ~\forall i \,.
\end{eqnarray*}
Now remember that $\lambda_{k}\neq0$ is strictly larger than
all the other $\lambda_{i}$ with $i>k$  since the eigenvalues are assumed
to be ordered. In addition, $d_{kk}^{\left(i\right)}\geq0\ ~\forall i$
and at least one $d_{kk}^{\left(i\right)}$ with $i\ge k$ must be nonzero,
otherwise the $d_{kk}^{\left(i\right)}$ would not sum up to $1$. Then
\begin{equation*}
\sum_{i\geq k}\lambda_{i}d_{kk}^{\left(i\right)}
\leq \lambda_{k}\sum_{i\geq k}d_{kk}^{\left(i\right)}
=\lambda_{k} \,,
\end{equation*}
with equality if and only if $d_{kk}^{\left(i\right)}=0$ for $i\neq
k$. In fact, equality has to hold since otherwise we would contradict
Eq.~\eqref{eq:secondequation}. We conclude that
\begin{equation*}
d_{kk}^{\left(i\right)}
=\vec{x}_{k}\cdot\mathcal{D}\left(P'_{i}\right)\cdot\vec{x}_{k}=\delta_{ik} \,.
\end{equation*}
Since $\vec{x}_{k}$ is normalized and arbitrary as long
as it lies in the eigenspace $\tilde{\mathcal E}_{k}$
of $\tilde{P}'_{k}$,
$\vec{x}_{k}$ must be an eigenvector of $\mathcal{D}\left(P'_{k}\right)$
with eigenvalue $1$. Consequently, the operator
$\mathcal{D}\left(P'_{k}\right)$ maps the eigenspace of
$\tilde{P}'_{k}$ onto itself. Now we are almost done
with showing that $\mathcal{D}\left(P'_{k}\right)$ and $\tilde{P}'_{k}$ are indeed
identical. Since $\tilde{\mathcal E}_{k}$ is mapped by
$\mathcal D$ into itself,
$\mathcal{D}\left(P'_{k}\right)$ has at least $\dim(\tilde{\mathcal E}_{k})$
eigenvalues equal to $1$. The fact that
$\mathcal{D}\left(P'_{k}\right)$ has exactly $\dim(\tilde{\mathcal E}_{k})$
eigenvalues equal to $1$ follows from $\mathcal{D}$ being
trace-preserving:
$\text{Tr}\left[\mathcal{D}\left(P'_{k}\right)\right]
=\text{Tr}\left[P'_{k}\right]=\dim (\mathcal E_{k})$
and $\dim (\mathcal E_{k})=\dim(\tilde{\mathcal E}_{k})$,
where $\text{Tr}\left[\mathcal{D}\left(P'_{k}\right)\right]$
is the sum over the eigenvalues of $\mathcal{D}\left(P'_{k}\right)$.
Since all eigenvalues of $\mathcal{D}\left(P'_{k}\right)$
are non-negative, all other eigenvalues must vanish.
Hence $\mathcal{D}\left(P'_{k}\right)=\tilde{P}'_{k}$.
This completes the induction and concludes the first step of the
$\Longleftarrow$ proof, i.e., we have shown that a unital dynamical
map that leaves the spectrum of a given arbitrary density matrix
invariant, maps projectors onto the eigenspaces of this density
matrix onto projectors of the same rank. This is specifically true
for one-dimensional projectors.
Iterating the argument for all density matrices in the complete set
and selecting a set $\mathcal{P}$ of $N$ orthogonal, one-dimensional
projectors,  it follows that these projectors will be mapped by
$\mathcal{D}$ onto another set of one-dimensional projectors.

In the second step of the $\Longleftarrow$ proof we still need to show
that the mapped set is also orthogonal. We denote the
complete set of projectors by $\{P_{i}\}$. From the first step of the
$\Longleftarrow$ proof we know that the $\tilde{P}_{i}$,
\begin{equation*}
\mathcal{D}\left(P_{i}\right)=\tilde{P}_{i} \,,
\end{equation*}
need to be one-dimensional projectors.
Using the unitality of $\mathcal{D}$, we see that
\begin{equation*}
\openone=\mathcal{D}\left(\openone\right)=\mathcal{D}\left(\sum_{i}P_{i}\right)=
\sum_{i}\mathcal{D}\left(P_{i}\right)=\sum_{i}\tilde{P}_{i} \,.
\end{equation*}
The unit matrix can only be summed by $N$ one-dimensional projectors
if these are orthogonal. Hence we have accomplished the second step, and
the lemma follows.

\textbf{Proof of Theorem 1}:
The equivalence relation of statement $(1) \Longleftrightarrow (2)$ has already
been proven in Ref.~\cite{ReichKochPRA13}. To complete
the proof of the more general Theorem~1, we are left with proving
$(2) \Longleftrightarrow (3)$.

$\left(2\right)\Longrightarrow\left(3\right)$:
If $\mathcal D$ maps a set of $N$ one-dimensional orthogonal
projectors onto another set of $N$ one-dimensional orthogonal
projectors, it leaves the spectrum of the projectors invariant.
This can be seen as follows. Projectors are idempotent and positive
semi-definite, hence their spectrum can only consist of zeros and
ones. Since the projector is one-dimensional,
its image under $\mathcal{D}$ has to be one-dimensional, too, and
there can only be one eigenvalue equal to one. Thus any
one-dimensional projector has the spectrum $\{1,0,0,\dots\}$ which
must be invariant under a mapping between one-dimensional orthogonal
projectors.
We now use the linearity of dynamical maps to show
that $\mathcal D$ must be unital.
Specifically, let $\{P_i\}$ be the initial set of orthogonal projectors
that is mapped to another set of orthogonal projectors, $\{\bar{P}_i\}$.
We find for the image of the totally mixed state,
$\rho_M = \frac{1}{N} \openone$,
\begin{eqnarray*}
  \mathcal{D}(\rho_M) & = &
  \mathcal{D}\left(\frac{1}{N} \sum_{i=1}^N P_i\right) =
  \frac{1}{N} \sum_{i=1}^N \mathcal{D}(P_i) \\
  & = & \frac{1}{N} \sum_{i=1}^N \bar{P}_i = \rho_M \,,
\end{eqnarray*}
i.e., $\mathcal D$ maps identity onto itself, making it unital. We can
thus use Lemma~1 to obtain that the
spectrum of a complete set of density matrices is invariant under
$\mathcal D$.
We now just have to add $\rho_{TR}=P_{TR}$ to realize a complete and
totally rotating set. The spectrum of $\rho_{TR}=P_{TR}$ is also
invariant under $\mathcal D$ since it is a one-dimensional projector
that is mapped onto another one-dimensional projector.

$\left(3\right)\Longrightarrow\left(2\right)$:
From Lemma~1, we obtain  that $\mathcal{D}$ maps a set of $N$
one-dimensional, orthogonal projectors onto another set of $N$
one-dimensional orthogonal projectors. We are thus only left with
showing that $\mathcal{D}$ maps a totally rotated projector onto a
one-dimensional projector: There always exists a density
matrix with a one-dimensional eigenspace corresponding to a totally
rotated projector $P_{TR}$ whose spectrum is invariant under the action
of $\mathcal{D}$. In the proof of Lemma~1, we have shown that
a dynamical map that leaves the spectrum of projectors invariant
maps these projectors onto projectors of the same rank.
Repeating the steps of the proof of Lemma~1, we see that the image of
$P_{TR}$ has to be a one-dimensional projector. This completes the
proof of Theorem~1.

\section{Adjoint of the Liouvillian in Lindblad form}%
\label{sec:erratum}

In general, for an optimization with Krotov's method where the forward
propagation takes the form
\begin{equation}
  \frac{d\Op\rho}{dt} = \mathcal{L}(\Op\rho)\,,
  \label{eq:forward_start}
\end{equation}
cf.~Eq.~\eqref{eq:LvN}, the equation of motion for the
backward-propagation is~\cite{ReichKochJCP12,BasilewitschAQT2019,GoerzSPP2019}
\begin{equation}
  \frac{d\Op\sigma}{dt} = -\mathcal{L}^\dagger(\Op\sigma)\,.
  \label{eq:backward_start}
\end{equation}
For numerical efficiency, Eq.~\eqref{eq:forward_start} is often evaluated \emph{directly}, by vectorizing the density matrices $\Op\rho$ and $\Op\sigma$, and applying the super-operator $\mathcal{L}$ as a (sparse) matrix, using matrix-vector multiplication.
The matrix representation of $\mathcal{L}^\dagger$ is then simply the adjoint of the matrix representation of $\mathcal{L}$.
This is in fact the approach we took for the numerical results shown in this paper.

A complication arises only when further writing out Eq.~\eqref{eq:forward_start} in Lindblad form,
\begin{subequations}
\begin{align}
\frac{d}{dt}\Op{\rho}(t)
    & =-i\left[\Op{H},\Op{\rho}(t)\right]+\mathcal{L}_{D}(\Op{\rho}(t))\label{eq:forward_intermediate}\\
    & =-i\left[\Op{H},\Op{\rho}(t)\right]+\sum_{k}\left(\Op{A}_{k}\Op{\rho}\Op{A}_{k}^{\dagger}-\frac{1}{2}\Op{A}_{k}^{\dagger}\Op{A}_{k}\Op{\rho}-\frac{1}{2}\Op{\rho}\Op{A}_{k}^{\dagger}\Op{A}_{k}\right)\,,\label{eq:forward_final}
\end{align}
\end{subequations}
with the Hamiltonian $\Op{H}$, the dissipator $\mathcal{L}_{D}$, and the Lindblad operators $\{\Op{A}_k\}$, prompting us to write Eq.~\eqref{eq:backward_start} in the corresponding form,
\begin{equation}
  \frac{d\Op\sigma}{dt}
  = -i[\Op H,\Op\sigma] -\mathcal{L}_D^\dagger(\Op\sigma)\,,
  \label{eq:backward_sigma}
\end{equation}
cf.~Eq.~\eqref{eq:backward}, where
\begin{equation}
  \mathcal{L}_{\mathcal{D}}^{\dagger}(\Op{\rho})
  =\sum_{k}\left(\Op{A}_{k}^{\dagger}\Op{\rho}\Op{A}_{k}-\frac{1}{2}\Op{A}_{k}^{\dagger}\Op{A}_{k}\Op{\rho}-\frac{1}{2}\Op{\rho}\Op{A}_{k}^{\dagger}\Op{A}_{k}\right)\,.
  \label{eq:L_D_dagger}
\end{equation}

When deriving an explicit form for the adjoint of the Lindbladian, Eqs.~(\ref{eq:backward_sigma},~\ref{eq:L_D_dagger}), it is important to note that $\mathcal{L}^\dagger(\Op{\rho}) \neq (\mathcal{L}(\Op{\rho}))^\dagger$.
Thus, the naive approach of simply taking the adjoint of the Liouville-von-Neumann equation, the right-hand-side of Eq.~\eqref{eq:forward_final}, would lead to a wrong result.
In the following, we derive the \emph{correct} Eqs.~(\ref{eq:backward_sigma},~\ref{eq:L_D_dagger}) from two different perspectives; first by formulating an abstract condition for the adjoint of the Liouvillian, and second by constructing the Liouville super-operator as a matrix.

\subsection{Adjoints of linear operators}

The most rigorous derivation of Eqs.~(\ref{eq:backward_sigma},~\ref{eq:L_D_dagger}) is based directly on the mathematical definition of an adjoint operator. To this end, we consider the density matrices as elements of a complete inner product space (a Hilbert space in the mathematical sense), using the Hilbert-Schmidt inner product
\begin{equation}
  \left\langle \Op{\rho}_{1},\Op{\rho}_{2}\right\rangle
  =\Tr[ \Op{\rho}_{1}^{\dagger}\Op{\rho}_{2}]\,,
\end{equation}
and the Liouvillian as a linear operator acting on that space.
For an arbitrary Hilbert space $\mathcal{H}$ the adjoint of an operator $\Op{O}$
is defined implicitly by the equation~\cite[p. 128]{SchaeferWolff1999}
\begin{equation}
  \forall x_{1},x_{2} \in \mathcal{H}:
  \langle x_{1}, \Op{O} x_{2}\rangle_{\mathcal{H}}
  = \langle \Op{O}^{\dagger} x_{1}, x_{2}\rangle_{\mathcal{H}}\,,
\end{equation}
where $\langle \cdot,\cdot\rangle_{\mathcal{H}}$ is the
scalar product associated with the Hilbert space. Thus, for the adjoint of the
Liouvillian,
\begin{equation}
  \forall\Op{\rho}_{1},\Op{\rho}_{2}:\
  \langle \Op{\rho}_{1},\mathcal{L}\Op{\rho}_{2}\rangle
  =\langle \mathcal{L}^{\dagger}\Op{\rho}_{1},\Op{\rho}_{2}\rangle\,.
\end{equation}

We use this rule to first convert Eq.~(\ref{eq:backward_start}) into the
form of Eq.~(\ref{eq:forward_intermediate}) by looking at an arbitrary
scalar product of the form $\left\langle \Op{\rho}_{1},\mathcal{L}\Op{\rho}_{2}\right\rangle $,
\begin{equation}
\begin{split}
  \left\langle \Op{\rho}_{1},\mathcal{L}\Op{\rho}_{2}\right\rangle
  & =\left\langle \Op{\rho}_{1},-i[\Op{H},\Op{\rho}_{2}]+\mathcal{L}_{D}\Op{\rho}_{2}\right\rangle \\
  & =\Tr[ -i\Op{\rho}_{1}^{\dagger}\Op{H}\Op{\rho}_{2}+i\Op{\rho}_{1}^{\dagger}\Op{\rho}_{2}\Op{H}] +\langle \Op{\rho}_{1},\mathcal{L}_{D}\Op{\rho}_{2}\rangle \\
  & =\Tr[ -i\Op{\rho}_{1}^{\dagger}\Op{H}\Op{\rho}_{2}+i\Op{H}\Op{\rho}_{1}^{\dagger}\Op{\rho}_{2}] +\langle \mathcal{L}_{D}^{\dagger}\Op{\rho}_{1},\Op{\rho}_{2}\rangle \\
  & =\Tr\left[ \left(i\Op{H}^{\dagger}\Op{\rho}_{1}\right)^{\dagger}\Op{\rho}_{2}+\left(-i\Op{\rho}_{1}\Op{H}^{\dagger}\right)^{\dagger}\Op{\rho}_{2}\right] +\langle \mathcal{L}_{D}^{\dagger}\Op{\rho}_{1},\Op{\rho}_{2}\rangle \\
  & =\left\langle i[\Op{H}^{\dagger},\Op{\rho}_{1}]+\mathcal{L}_{D}^{\dagger}\Op{\rho}_{1},\Op{\rho}_{2}\right\rangle \equiv \left\langle \mathcal{L}^{\dagger}\Op{\rho}_{1},\Op{\rho}_{2}\right\rangle ,
\end{split}
\end{equation}
from which the definition of the adjoint Liouvillian follows as
\begin{equation}
  \mathcal{L}^{\dagger}\Op{\rho}
  =i [\Op{H}^{\dagger},\Op{\rho}]+\mathcal{L}_{D}^{\dagger}\Op{\rho}\,.
\end{equation}
Since we assume the Hamiltonian $\Op{H}$ to be Hermitian, applying this
to the equation of motion~\eqref{eq:backward_start} for the co-state in Krotov's
method, we find
\begin{equation}
  \frac{d}{dt}\Op{\sigma}(t)
  =-\mathcal{L}^{\dagger}\Op{\sigma}(t)
  =-i [\Op{H},\Op{\sigma}(t)]-\mathcal{L}_{D}^{\dagger}\Op{\sigma}(t)\,,
\end{equation}
cf.\ Eq.~\eqref{eq:backward_sigma}.

We now continue to write out $\mathcal{L}_{D}^{\dagger}\Op{\sigma}(t)$ in terms
of the set of Lindblad operators $\{ \Op{A}_{k}\}$. Once again, we
start with the general form of a Liouville-space scalar product involving
$\mathcal{L}_{D}$,
\begin{equation}
\begin{split}
 \left\langle \Op{\rho}_{1},\mathcal{L}_{D}\Op{\rho}_{2}\right\rangle
  & =\left\langle \Op{\rho}_{1},\sum_{k}\left(\Op{A}_{k}\Op{\rho}_{2}\Op{A}_{k}^{\dagger}-\frac{1}{2}\Op{A}_{k}^{\dagger}\Op{A}_{k}\Op{\rho}_{2}-\frac{1}{2}\Op{\rho}_{2}\Op{A}_{k}^{\dagger}\Op{A}_{k}\right)\right\rangle \\
  & =\sum_{k}\Tr\left[ \Op{\rho}_{1}^{\dagger}\Op{A}_{k}\Op{\rho}_{2}\Op{A}_{k}^{\dagger}-\frac{1}{2}\Op{\rho}_{1}^{\dagger}\Op{A}_{k}^{\dagger}\Op{A}_{k}\Op{\rho}_{2}-\frac{1}{2}\Op{\rho}_{1}^{\dagger}\Op{\rho}_{2}\Op{A}_{k}^{\dagger}\Op{A}_{k}\right] \\
  & =\sum_{k}\Tr\left[ \Op{A}_{k}^{\dagger}\Op{\rho}_{1}^{\dagger}\Op{A}_{k}\Op{\rho}_{2}-\frac{1}{2}\Op{\rho}_{1}^{\dagger}\Op{A}_{k}^{\dagger}\Op{A}_{k}\Op{\rho}_{2}-\frac{1}{2}\Op{A}_{k}^{\dagger}\Op{A}_{k}\Op{\rho}_{1}^{\dagger}\Op{\rho}_{2}\right] \\
  & =\sum_{k}\Tr\left[ \left(\Op{A}_{k}^{\dagger}\Op{\rho}_{1}\Op{A}_{k}\right)^{\dagger}\Op{\rho}_{2}-\frac{1}{2}\left(\Op{A}_{k}^{\dagger}\Op{A}_{k}\Op{\rho}_{1}\right)^{\dagger}\Op{\rho}_{2}-\frac{1}{2}\left(\Op{\rho}_{1}\Op{A}_{k}^{\dagger}\Op{A}_{k}\right)^{\dagger}\Op{\rho}_{2}\right] \\
  & =\sum_{k}\left\langle \Op{A}_{k}^{\dagger}\Op{\rho}_{1}\Op{A}_{k}-\frac{1}{2}\Op{A}_{k}^{\dagger}\Op{A}_{k}\Op{\rho}_{1}-\frac{1}{2}\Op{\rho}_{1}\Op{A}_{k}^{\dagger}\Op{A}_{k},\Op{\rho}_{2}\right\rangle \\
  & =\left\langle \sum_{k}\left(\Op{A}_{k}^{\dagger}\Op{\rho}_{1}\Op{A}_{k}-\frac{1}{2}\Op{A}_{k}^{\dagger}\Op{A}_{k}\Op{\rho}_{1}-\frac{1}{2}\Op{\rho}_{1}\Op{A}_{k}^{\dagger}\Op{A}_{k}\right),\Op{\rho}_{2}\right\rangle \equiv\left\langle \mathcal{L}_{\mathcal{D}}^{\dagger}\Op{\rho}_{1},\Op{\rho}_{2}\right\rangle \,,
\end{split}
\end{equation}
from which we can read off Eq.~\eqref{eq:L_D_dagger}.

\subsection{Vectorization of Liouville space}

Alternatively, Eqs.~(\ref{eq:backward_sigma},~\ref{eq:L_D_dagger}) can also be derived by making the vectorization of density matrices and the subsequent treatment of the Liouvillian as a sparse super-operator matrix more explicit. While this approach is less elegant than the abstract mathematical derivation, it provides the exact construction of the Liouville super-operator matrix from the Hamiltonian and the set of Lindblad operators, which is immensely useful for numerical applications.

First, we vectorize $\Op{\rho}$ with elements $\rho_{ij}$ into a column vector
$\vec{\rho} = \vectorize(\Op{\rho})$ with elements $\rho_{(ij)} = \rho_{ij}$,
using a double-index $(ij)$ where $i$ varies more quickly than $j$ (column-major
order).  We can then show element-wise that
\begin{equation}
  \vectorize\left(\Op{A} \Op{\rho} \Op{B} \right)
  = \left(\mat{B}^{T} \otimes \mat{A}\right) \vec{\rho}\,,
  \label{eq:A_rho_B_vec}
\end{equation}
where $\mat{A}$ and $\mat{B}$ are the matrix representations of $\Op{A}$ and
$\Op{B}$ and $\mat{B}^T$ is the transpose of $\mat{B}$.
For the left-hand-side, we find that
\begin{equation}
  \left(\Op{A} \Op{\rho} \Op{B}\right)_{ij}
  = \sum_{k'k} A_{ik'} \rho_{k'k} B_{kj}\,.
  \label{eq:A_rho_B_vec_lhs}
\end{equation}
On the right-hand-side, if we define $\mat{C} = \mat{B}^T \otimes \mat{A}$ and
label the entries of $\mat{C}$ with double-indices, then the elements of the
matrix-vector product with $\vec{\rho}$ are (double-indexed)
\begin{equation}
  (\mat{C} \vec{\rho}\,)_{(ij)} = \sum_{(i'j')} C_{(ij)(i'j)} \rho_{(i'j')}\,.
\end{equation}
The double-indices for $\mat{C}$ must match those for $\vec{\rho}$.
To match the proper column-major ordering, we identify
the elements of a general tensor product as~\footnote{This can be seen, e.g., by
writing out the tensor product of two $2 \times 2$ matrices and comparing
indices with the proper double-index labels.}
\begin{equation}
  (\mat{A} \otimes \mat{B})_{(ij)(i'j')} = A_{jj'} B_{ii'}\,,
\end{equation}
or for the specific tensor product $\mat{C} = \mat{B}^T \otimes \mat{A}$,
\begin{equation}
  \mat{C}_{(ij)(i'j')}
  = (\mat{B}^T \otimes \mat{A})_{(ij)(i'j')}
  = B_{j'j} A_{ii'}\,.
\end{equation}
Thus, for the right-hand-side of Eq.~\eqref{eq:A_rho_B_vec}, we have
\begin{equation}
\begin{split}
  \left[\left(\mat{B}^{T} \otimes \mat{A}\right) \vec{\rho}\,\right]_{(ij)}
  &= \sum_{i'j'} B_{j'j} A_{ii'} \rho_{i'j'} \qquad \text{\scriptsize ($i'\!\rightarrow\! k'$, $j' \!\rightarrow\! k )$}\\
  &= \sum_{k'k} A_{ik'} \rho_{k'k} B_{kj}\,.
\end{split}
\end{equation}
This matches Eq.~\eqref{eq:A_rho_B_vec_lhs} and thus proves
Eq.~\eqref{eq:A_rho_B_vec}.

We can now use Eq.~\eqref{eq:A_rho_B_vec} to write out the matrix representation
$\mat{L}$ of the Liouvillian $\mathcal{L}$,
\begin{equation}
  \mat{L} =
    -i (\identity \otimes \mat{H}) + i (\mat{H}^T \otimes \identity)
    + \sum_k \left[
      (\mat{A}_k^\dagger)^T \otimes \mat{A}_k
      - \half \left(\identity \otimes \mat{A}_k^\dagger \mat{A}_k\right)
      - \half \left((\mat{A}_k^\dagger \mat{A}_k)^T \otimes \identity\right)
      \right]\,.
  \label{eq:L_superop_repr}
\end{equation}
To evaluate the right-hand-side of Eq.~\eqref{eq:backward_start},  we calculate
the matrix representation of $- \mathcal{L}^\dagger$ as
\begin{equation}
  -\mat{L}^\dagger =
    -i (\identity \otimes \mat{H}) + i (\mat{H}^T \otimes \identity)
    - \sum_k \left[
      \mat{A}_k^T \otimes \mat{A}_k^\dagger
      - \half \left(\identity \otimes \mat{A}_k^\dagger \mat{A}_k\right)
      - \half \left((\mat{A}_k^\dagger \mat{A}_k)^T \otimes \identity\right)
      \right]\,,
  \label{eq:L_dag_superop_repr}
\end{equation}
where we have used $(\mat{A} \otimes \mat{B})^\dagger = \mat{A}^\dagger \otimes
\mat{B}^\dagger$ and $(\mat{A}\mat{B})^\dagger = \mat{B}^\dagger
\mat{A}^\dagger$.  Compared to $\mat{L}$, only the sign of the sum and the first
term inside the sum change. We note that the super-operator $\mat{A}_k^T \otimes
\mat{A}_k^\dagger$  corresponds to $\Op{A}_k^\dagger \Op{\rho} \Op{A}_k$
according to Eq.~\eqref{eq:A_rho_B_vec}, and thus recover
Eq.~\eqref{eq:L_D_dagger}.

\bibliography{oct}

\end{document}